\documentclass[tradiabstract]{aa}

\usepackage{graphicx}
\usepackage{natbib}
\usepackage{scalerel}
\usepackage{capt-of}
\usepackage{float}
\usepackage[table]{xcolor}
\usepackage[pdfencoding=auto,psdextra]{hyperref}
\hypersetup{
    colorlinks=true,
    linkcolor=blue,
    filecolor=magenta,      
    urlcolor=blue,
    citecolor=blue
}
\makeatletter
\renewcommand*\aa@pageof{, page \thepage{} of \pageref*{LastPage}}
\makeatother

\usepackage[utf8]{inputenc}

\usepackage{placeins}
\usepackage{euclid}
\usepackage{orcidlink}
\let\orcid\orcidlink
\usepackage[nolist,nohyperlinks]{acronym}
\usepackage[capitalise]{cleveref}
\crefname{section}{Sect.}{Sects.}

\usepackage{longtable}
\usepackage[separate-uncertainty]{siunitx}

\newcommand{\rband}{\ensuremath{r}-band\xspace}
\newcommand{\uband}{\ensuremath{u}-band\xspace}
\newcommand{\specialcell}[2][l]{%
  \begin{tabular}[#1]{@{}l@{}}#2\end{tabular}}

\begin{document}
\acrodef{IAU}{International Astronomical Union}
\acrodef{IMF}{initial mass function}
\acrodef{PSF}{point spread function}
\acrodef{RMS}{root mean square}
\acrodef{SNR}[S/N]{signal-to-noise ratio}

\acrodef{CFIS}{the Canada-France Imaging Survey}
\acrodef{DES}{the Dark Energy Survey}
\acrodef{DESI}{the Dark Energy Spectroscopic Instrument}
\acrodef{HSC}{the Subaru Hyper Suprime-Cam}
\acrodef{KiDS}{the Kilo-Degree Survey}
\acrodef{Pan-STARRS}{the Panoramic Survey Telescope And Rapid Response System}
\acrodef{SDSS}{the Sloan Digital Sky Survey}
\acrodef{UNIONS}{the Ultraviolet Near Infrared Optical Northern Survey}
\acrodef{WHIGS}{the Waterloo Hawaii Institute for Astronomy G-band Survey}
\acrodef{WISHES}{the Wide Imaging with Subaru Hyper Suprime-Cam of the Euclid Sky}

\acrodef{CFHT}{the Canada-France-Hawaii Telescope}
\acrodef{NOT}{the Nordic Optical Telescope}
\acrodef{ALFOSC}{Alhambra Faint Object Spectrograph and Camera}

\newacro{SLED}{the Strong Lens Database}
\acrodef{SIE}{single isothermal ellipsoid}
\acrodefplural{SIE}{single isothermal ellipsoids}

\acrodef{CNN}{convolutional neural network}
\acrodefplural{CNN}{convolutional neural networks}
\acrodef{ResNet}{residual neural network}
\acrodef{TPR}{true positive rate}
\acrodef{FPR}{false positive rate}
\acrodef{ROC}{receiver operating characteristic}

\acrodef{SPS}{stellar population synthesis}
\acrodef{LRG}{luminous red galaxy}
\acrodefplural{LRG}{luminous red galaxies}


\acrodef{MAD}{median absolute deviation}

   \title{Searching for strong lensing by late-type galaxies in UNIONS}

\author{J.~A.~Acevedo~Barroso\orcid{0000-0002-9654-1711}\thanks{\email{javier.acevedobarroso@gmail.com}}
          \inst{1}
          \and
          B.~Clément\orcid{0000-0002-7966-3661}\inst{1, 2}
          \and
          F.~Courbin\orcid{0000-0003-0758-6510}\inst{1, 3, 4}
          \and
          R.~Gavazzi\orcid{0000-0002-5540-6935}\inst{5, 6}
          \and
          C.~Lemon\orcid{0000-0003-2456-9317}\inst{1, 7}
          \and
          K.~Rojas\orcid{0000-0003-1391-6854}\inst{8, 9}
          \and
          D.~Scott\orcid{0000-0002-6878-9840}\inst{10} 
          \and
          S.~Gwyn\orcid{0000-0001-8221-8406}\inst{11}
          \and
          F.~Hammer\orcid{0000-0002-2165-5044}\inst{12} 
          \and
          M.~J.~Hudson\orcid{0000-0002-1437-3786}\inst{13, 14, 15} 
          \and
          E.~A.~Magnier\orcid{0000-0002-7965-2815}\inst{16} 
          }

   \institute{Institute of Physics, Laboratory of Astrophysics, Ecole Polytechnique Fédérale de Lausanne (EPFL), Observatoire de Sauverny, 1290 Versoix, Switzerland
         \and
         SCITAS, Ecole Polytechnique F\'ed\'erale de Lausanne (EPFL), 1015 Lausanne, Switzerland
         \and
        Institut de Ci\`{e}ncies del Cosmos (ICCUB), Universitat de Barcelona (IEEC-UB), Mart\'{i} i Franqu\`{e}s 1, 08028 Barcelona, Spain
        \and
        Instituci\'o Catalana de Recerca i Estudis Avan\c{c}ats (ICREA), Passeig de Llu\'{\i}s Companys 23, 08010 Barcelona, Spain
         \and
         Laboratoire d’Astrophysique de Marseille, UMR7326, Aix-Marseille Université, CNRS, CNES, 13013 Marseille, France
        \and
        Institut d'Astrophysique de Paris, UMR 7095, CNRS, and Sorbonne Universit\'e, 98 bis boulevard Arago, 75014 Paris, France
        \and
        Oskar Klein Centre, Department of Physics, Stockholm University, SE-106 91 Stockholm, Sweden
         \and
        University of Applied Sciences and Arts of Northwestern Switzerland, School of Engineering, 5210 Windisch, Switzerland
         \and
             Institute of Cosmology and Gravitation, University of Portsmouth, Burnaby Rd, Portsmouth PO1 3FX, UK
         \and
         Department of Physics and Astronomy, University of British Columbia, Vancouver, BC V6T 1Z1, Canada
         \and
         Canadian Astronomy Data Centre, Herzberg Astronomy and Astrophysics, National Research Council, 5071 West Saanich Rd Victoria BC V9E 2E7, Canada
         \and
         LIRA, Observatoire de Paris, Université PSL, CNRS, Place Jules Janssen 92195, Meudon, France 
         \and
         Department of Physics and Astronomy, University of Waterloo, Waterloo, ON, N2L 3G1, Canada
         \and
         Waterloo Centre for Astrophysics, Waterloo, ON, N2L 3G1, Canada
         \and
         Perimeter Institute for Theoretical Physics, 31 Caroline St. N., Waterloo, ON, N2L 2Y5, Canada
         \and
         Institute for Astronomy, University of Hawaii, 2680 Woodlawn Drive, Honolulu, HI 96822, USA
             }


 \abstract{
Recent wide-field galaxy surveys have led to an explosion in the number of galaxy-scale strong gravitational lens candidates.
However, the vast majority of them feature massive luminous red galaxies as the main deflectors, with late-type galaxies being vastly under-represented.
This work presents a dedicated search for lensing by edge-on late-type galaxies in the Ultraviolet Near Infrared Optical Northern Survey (UNIONS).
The search covers $3600\,\deg^2$ of $r$-band observations taken from the Canada-France-Hawaii Telescope.
We considered all sources with magnitudes in the range $17 < r < 20.5$, without any colour pre-selection, yielding a parent sample of seven million sources.
We characterised our parent sample via the visual inspection of $120\,000$ sources selected at random.
From it, we estimate, with a 68\% confidence interval, that 1 in every $30\,000$ sources is an edge-on lens candidate, with at least eight high-quality candidates in the parent sample.
This corresponds to one candidate per $17\,000$ edge-on late-type galaxies.
Our search relied on a convolutional neural network (CNN) to select a reduced sample of candidates, which we followed with a visual inspection to curate the final sample.
The CNN was trained from scratch using simulated $r$-band observations of edge-on lenses, and real observations of non-lenses.
We found 61 good edge-on lens candidates using the CNN.
Moreover, combining the CNN candidates with those found serendipitously and those identified while characterising the parent sample, we discovered 4 grade A, 20 grade B, and 58 grade C edge-on lens candidates, effectively doubling the known sample of these systems.
We also discovered 16 grade A, 16 grade B, and 18 grade C lens candidates of other types.
Finally, based on the characterisation of the parent sample, we estimate that our search found around 60\% of the bright grade A and B edge-on lens candidates within the parent sample.

 }


   \maketitle

\section{Introduction}
Strong gravitational lensing has become an essential tool for probing the extragalactic Universe.
Its applications range from the detection and study of high-redshift sources \citep[e.g.,][]{2023MNRAS.524.5486A,2023ApJ...955...13B}, including the mapping of the mass distributions responsible for the lensing effect \citep[e.g.,][]{2019MNRAS.483.5649S,2022MNRAS.517.3275E,2023ApJ...945...49M}, to the constraining of cosmological parameters, most notably $H_0$ when the lensed source exhibits time variability \citep[e.g.,][]{2020MNRAS.498.1420W, 2020A&A...643A.165B,2023Sci...380.1322K,2024ApJ...970..102C}.
Additionally, the analysis of flux ratios between multiply lensed images, alongside small-scale distortions in the lensed arcs,
allows for the detection of mass clumps in the line of sight \citep[e.g.,][]{2010MNRAS.408.1969V,2023MNRAS.521.3298N,2024MNRAS.530.2960N,2024MNRAS.533.1687G,2024MNRAS.52710480N}.
Furthermore, combining lens modelling with a second probe of mass can break degeneracies in mass models that limit single-probe analyses, such as the baryonic and dark matter degeneracy \citep[e.g.,][]{2010ApJ...724..511A,2019A&A...631A..40S,2025MNRAS.541....1S}, the bulge-disc degeneracy for late-type galaxies \citep[e.g.,][]{2011MNRAS.417.1621D,2012ApJ...750...10S}, or the mass-sheet degeneracy, which affects lens models in general \citep[e.g.,][]{2023A&A...673A...9S,2024MNRAS.533..795K}.
In particular, galaxy-scale lensing, when combined with deep spectroscopic observations or multi-band photometry, can probe the stellar \ac{IMF} of the lensing galaxies
\citep[see,][]{2010MNRAS.409L..30F, 2012MNRAS.419..656C, 2014MNRAS.438.3594D,2014MNRAS.437.1950B,2019A&A...630A..71S}.

However, these studies are limited by their sample size, and their results can change with improved samples and statistics.
This is especially the case for late-type galaxies, for which the number of confirmed lenses amounts to a few dozen objects.
Whether the \ac{IMF} is the same for late-type galaxies as it is for early-type galaxies, in what way the \ac{IMF} might differ in the bulge from the disc for a given galaxy, and the universality of the \ac{IMF} for late-type galaxies, are all open questions that can be tackled with large samples of late-type lens galaxies.
In this context, the Sloan WFC Edge-on Late-type Lens Survey (SWELLS; WFC – wide field camera) launched a campaign in the early 2010s to find and study lensing by edge-on late-type galaxies (edge-on lenses) while trying to answer the aforementioned questions.
They produced a series of six articles covering the curation and characterisation of a spectroscopically selected sample of edge-on lenses \citep{2011MNRAS.417.1601T}; the combination of lens modelling with spatially resolved kinematics to break the disc-halo degeneracy and probing of the \ac{IMF} for a single galaxy \citep{2011MNRAS.417.1621D, 2012MNRAS.423.1073B}; the systematic comparison between the total mass inside the Einstein radius and the stellar mass estimated from \ac{SPS} models to study the \ac{IMF} for medium-sized samples ($\approx\,20$) of edge-on lenses \citep{2012MNRAS.422.3574B,2014MNRAS.437.1950B}; 
and the study of the \ac{IMF} normalisation of the bulge and the disc independently for a sample of five edge-on lenses using lens modelling, kinematics, and \ac{SPS} \citep{2013MNRAS.428.3183D}.
To date, the SWELLS papers represent the largest study of edge-on lenses, but it is still limited to a few dozen systems.

Galaxy-scale strong lensing events are very rare.
The likelihood of observing such events depends on the mass distribution of the deflectors and the redshift distributions of foreground lenses and background sources.
As a result, only about one source in thousands exhibits lensed images \citep[e.g.,][]{2010MNRAS.405.2579O, 2015ApJ...811...20C,2024MNRAS.532.1832F,2025A&A...697A..14A}, and these sources are not easy to see in ground-based imaging, because of the small image separation on the plane of the sky and because of the low luminosity contrast between the lens light and the background source.
This is true when the lens galaxy is a massive \ac{LRG}, but the situation is even worse for the less massive late-type galaxies.
In fact, only about a hundred lens candidates have a late-type deflector, out of approximately twenty thousand lens candidates known by the community (see Vernardos et al., in prep.).
Moreover, even though the number of lens candidates is growing orders of magnitude faster since the adoption of machine learning in lens finding, the number of late-type candidates has been growing at a much more modest rate, hinting at the necessity of lens searches that target these objects directly.

Recent massive optical lens searches have targeted wide-sky surveys to maximise the volume probed while also using neural networks to select samples of potential lens candidates.
These surveys include
\acl{DES}\acused{DES} \citep[DES;][]{2019ApJS..243...17J,2019MNRAS.484.5330J,2022A&A...668A..73R,2022ApJS..259...27O},
\acl{KiDS}\acused{KiDS} \citep[KiDS;][]{2017MNRAS.472.1129P,2019MNRAS.484.3879P,2020MNRAS.497..556H,2020ApJ...899...30L,2021ApJ...923...16L},
\acl{HSC}\acused{HSC} \citep[HSC;][]{2021A&A...653L...6C,2022A&A...662A...4S,2024MNRAS.535.1625J,2025A&A...693A.291S},
\ac{DESI} Legacy Imaging Survey \citep{2020ApJ...894...78H,2021ApJ...909...27H,2022ApJ...932..107S,2024ApJS..274...16S},
the \Euclid Early Release Observations \citep{Nagam25},
and \acl{UNIONS}\acused{UNIONS} \citep[UNIONS;][]{2022A&A...666A...1S}.
Most of the analyses rely on \acp{CNN} as the main classifier, but visual transformers are also being explored as an alternative \citep{2024A&A...688A..34G,2025arXiv250115679G}.

Given the extremely low prevalence of lensing events, it is common to restrict lens searches to the most massive galaxies, often making a colour pre-selection for \acp{LRG}.
These pre-selections remove most of the late-type galaxy lenses, excluding them from the lens search even before the \acp{CNN} inspect them.
However, even if there is no colour pre-selection, there are still not enough confirmed lenses to train neural networks using only real observations.
Instead, the \acp{CNN} are trained on mock observations simulating the lensing effect.
These mock lenses are being refined, from the first ones by \citet{2017MNRAS.472.1129P} randomly adding fully simulated lensed sources to \acp{LRG} to the recent use of real deep \textit{HST} observations for the potential sources and the dynamics of the potential deflectors to select a reasonable lensing mass \citep{2020A&A...644A.163C,2022A&A...668A..73R,2022A&A...666A...1S}.
The problem with this methodology is that it limits the search to the lens candidates that look the most similar to the mock lenses used to train the networks.
And since these studies mostly use \acp{LRG} for the deflector galaxies, the late-type lenses are also indirectly excluded, even if they are part of the parent sample inspected by the networks.

Another limitation in the discovery of late-type galaxy lenses is their relatively lower mass compared to early-type galaxies.
This, combined with the characteristic bulge-disc morphology of late-type galaxies, results in most of the lensing being caused by the bulge when the galaxy is not viewed edge-on.
In such cases, the Einstein radius is often too small to be resolved by most ground-based surveys.
Consequently, it is expected that most late-type galaxy lenses detectable by ground-based surveys will have an edge-on orientation.

In this work, we search for galaxy-scale lensing by late-type edge-on disc galaxies (hereafter `edge-on lenses') in \ac{CFIS}, now a part of \ac{UNIONS}.
We chose this survey because of its excellent image quality,
large area coverage,
and its complete overlap with the on-going Euclid Wide Survey \citep{Scaramella-EP1,EuclidSkyOverview}.
This work represents the second lens search specifically targeting edge-on lenses in optical data, after \citet{2010A&A...517A..25S}, who also used data from \ac{CFHT}.
However, this is the first time it is being done using machine learning and the methodology popularised in the past decade.
Our main objective is to substantially increase the sample of known edge-on lens candidates, while also exploring why they are not being found by the recent massive lens searches.
Additionally, we aim to estimate the prevalence and total number of edge-on lenses in the data and to develop a framework to evaluate the performance of our CNN search without relying on simulated lenses.

This paper is organised as follows.
We present the UNIONS-CFIS observations and the selection of the parent sample in \cref{sect:data}.
Then, in \cref{sect: estimating the number of lenses (prevalence study)}, we characterise the parent sample by estimating the number of lenses in it.
We describe the process to simulate mock observations of edge-on lenses in \cref{sect:simulations},
and present the training and deployment of the \ac{CNN} in \cref{sect: lens finding with CMU DeepLens}.
We present spectroscopic follow-up of two candidates from \ac{NOT} in \cref{sect:spectroscopy}.
Finally, we discuss our findings and conclude in \cref{sect:discussion,sect:conclusions}.

\section{Data}
\label{sect:data}

We used \textit{r}-band data from \ac{CFIS} data release 3, which was conducted using the 3.6-m \ac{CFHT} on Maunakea, Hawaii.
CFIS provides deep \textit{r}- and \textit{u}-band imaging to \ac{UNIONS}, a collaboration of wide-field imaging surveys of the northern sky involving CFIS, as well as members of \ac{Pan-STARRS} team, \ac{WHIGS} team, and \ac{WISHES} team.
Deep \textit{i}-band and moderately deep \textit{z}-band imaging data are provided by \ac{Pan-STARRS}, deep \textit{g}-band data are provided by WHIGS, and deep \textit{z}-band data are provided by the WISHES team.
\ac{UNIONS} aims to provide ground-based complementary data for the \Euclid mission, as well as maximising the science output of these ground-based large-area deep imaging surveys \citep{2017ApJ...848..128I, 2022A&A...666A.162G}.

The \ac{CFIS} data release 3 covers $\num{7000}\,\rm{deg}^2$ down to a median depth of $24.3$ in the \textit{u}-band, and $\num{3600}\,\rm{deg}^2$ down to a median depth of $25.2$ in the \textit{r}-band.
Both depths are for point sources with $5\,\sigma$ detection in a $\ang{;;2}$ aperture.
The median seeing of the data is \mbox{\ang{;;0.8}} and \mbox{\ang{;;0.6}} for the \textit{u}- and \textit{r}-bands, respectively.
The individual observations were reduced, calibrated and co-added at the Canadian Astronomy Data Centre (CADC) using a modified version of the \texttt{MegaPipe} pipeline \citep{2008PASP..120..212G, 2019ASPC..523..649G}.
The final images have a pixel size of \mbox{\ang{;;0.1857}}.
Additionally, models of the \ac{PSF} and its spatial variations were produced for every co-added frame using \texttt{PSFEx} \citep{2011ASPC..442..435B}. 
In the present work, we focus on CFIS \textit{r}-band data due to their exquisite seeing quality.
This is crucial for our specific search, since late-type edge-on galaxies have a smaller lensing cross-section than the typical \acp{LRG} used in other searches.
As a consequence, the potential lensed images are more often hidden in the glare of the lens light than for systems with early-type lens galaxies. 

We classify detections with flux radius larger than one standard deviation above the median value as extended sources.
We then extract postage stamps of $66 \times 66$ pixels ($\ang{;;12.256}\times\ang{;;12.256}$) for every extended source within the range $17.0< r < 20.5$, amounting to \num{6978977} stamps.
For each source, we also use an \ac{RMS} map and a \ac{PSF} image generated from the \texttt{PSFEx} model and oversampled by a factor of 2.
This is our `parent sample', which is further characterised in \cref{sect: estimating the number of lenses (prevalence study)}.

\section{Estimating the number of lenses in the parent sample}
\label{sect: estimating the number of lenses (prevalence study)}

\subsection{The challenges of finding rare objects}

We approach the lens search as a binary image classification problem, where the target class is composed of stamps displaying reasonable lensing features.
To tackle this problem, we use a supervised machine-learning approach, training our \ac{CNN} on a set of correctly classified images.
The goal is that, upon encountering a new image, the network will correctly classify it based on the patterns learned from the training data.
While this approach is standard practice in image classification, the extremely low prevalence of edge-on lenses on the sky presents additional challenges.

The first challenge is that there are not enough observations of edge-on lenses to train a CNN from scratch, which requires tens of thousands of examples, while only about a hundred edge-on lens candidates are currently known.
Consequently, we have to produce a simulated training set for the network.
It remains an open question how much the use of simulations affects the precision and completeness of the lens search.

The second challenge resides in a combination of networks producing false positives and having to face the so-called `base rate fallacy': even if the network correctly classifies non-lenses most of the time, for example $99.9\%$ of the sample, the remaining  misclassified $0.1\%$ is still large enough to dominate the sample of lens candidates suggested by the network.
This is best illustrated using the Bayes' rule, stating that
\begin{equation}
    \label{eq: bayes rule}
    P(\text{L}\, |\, \text{C}) = \frac{P(\text{C}\, |\, \text{L}) P(\text{L})}{P(\text{C})},
\end{equation}
where $P(\text{L}\, |\, \text{C})$ is the probability that something classified as a lens candidate is indeed a lens, this represents the ratio of lenses over machine-selected candidates, and is often called the `purity' of the lens search.
On the other hand, $P(\text{C}\, |\, \text{L})$ is the probability that the \ac{CNN} correctly classifies the image of a lens as a lens.
This is also known as the \ac{TPR}, or recall.
$P(\text{L})$ is the probability of a random stamp being a lens,
which is the prevalence of lensing on the sky.
$P(\text{C})$ is the probability that the \ac{CNN} classifies any given stamp as a lens. Since every stamp shows either a lens or not, we marginalised the last term as
\begin{equation}
    P(\text{C}) = P(\text{C}\, |\, \text{L})P(\text{L}) + P(\text{C}\, |\, \text{Non-L})(1-P(\text{L})),
\end{equation}
with $P(\text{C}\, |\, \text{Non-L})$ being the probability that the network mistakenly classifies an image that does not contain any lensing features as a lens.
This is known as the \ac{FPR}.
~\Cref{eq: bayes rule} becomes
\begin{equation}
\label{eq: purity}
    P(\text{L}\, |\, \text{C}) = \frac{P(\text{C}\, |\, \text{L}) P(\text{L})}{P(\text{C}\, |\, \text{L})P(\text{L}) + P(\text{C}\, |\, \text{Non-L})(1-P(\text{L}))}\,.
\end{equation}
Taking the conservative estimate that one object shows lensing features for every \num{10000} galaxies, then $P(\text{L})=10^{-4}$. If we further assume a network capable of correctly classifying all lenses in a sample and 99\% of the non-lenses, then $P(\text{C}\, |\, \text{L})=1$ and $P(\text{C}\, |\, \text{Non-L})=0.01$.
From there, \cref{eq: purity} implies that only $1\%$ of the stamps classified as lenses by the network are in fact real lenses, that is, $P(\text{L}\, |\, \text{C})\approx0.01$.

The latter point is probably the main limitation of automated lens finding approaches in general, because it challenges even the very best networks currently available.
Moreover, this is further complicated by the fact that many objects that have nothing to do with lensing do display morphological structures that efficiently mimic lensing.
This is the third challenge to face. 
An immediate consequence of all the above is that even when using the best techniques, a final visual inspection step is needed to compile a reliable list of lens candidates. 

\subsection{The visual inspection}

To estimate the number of galaxy-scale lenses in the parent sample, we conduct a systematic visual inspection of randomly selected sources.
This procedure allows us to establish a baseline against which we can compare the performance of our \ac{CNN} classifier.
Additionally, it provides some estimates of the potential of the parent sample for future, more general, lens finding efforts.
For this, we split the parent sample into four \rband bins of magnitude $(17,18]$, $(18,19]$, $(19,20]$, and $(20,20.5]$.
We then randomly select \num{30000} sources per bin to be inspected, for a total of \num{120000} sources probed.
A similar procedure is used in \citet{2025A&A...697A..14A} to estimate the number of lens candidates in the entire Euclid Wide Survey.

For the visual inspection, we followed a three-step methodology in which the sources selected in one step go on to be reinspected in the next one:
\begin{enumerate}[Step 1:]
    \item The experts individually inspect the sources in mosaics,
    \item The experts individually inspect the sources one at a time, and
    \item The experts collectively inspect the sources one at the time.
\end{enumerate}
For this purpose, we reimplemented the visualisation tools used in \citet{2022A&A...668A..73R} and \citet{2022A&A...666A...1S} using the \texttt{Qt6} framework.\footnote{Available at \url{https://github.com/ClarkGuilty/Qt-stamp-visualizer}.}
The tools correspond to a mosaic viewer that shows multiple sources in a rectangular grid, and a one-by-one sequential tool that displays a single source at a time.
Both tools allow the user to change the colour-map and scaling function used to display the stamps.
The mosaic tool is used for binary classification: the user goes through pages of mosaics and clicks on the stamps that show hints of lensing.
By contrast, the one-by-one sequential tool allows for detailed classification of sources into the following non-overlapping categories:
\begin{itemize}
    \item A: The source shows clear lensing features and no additional information is needed.
    \item B: The source  shows lensing features but additional information is required to confirm its lensing nature.
    \item C: The source  shows lensing features, but these can be explained by other phenomena.
    \item X: The source is definitively not a lens.
\end{itemize}
Grade A candidates clearly display multiple images or lensed arcs following typical lensing configurations.
By contrast, grade B candidates may exhibit potential multiple images but lack typical lensing symmetries, usually spectroscopy or high-resolution imaging is enough to confirm these candidates lenses.
Grade C candidates still resemble simulated mock lenses, but the imaging can also be explained by random alignment of galaxies or a merger event.
Because of this, grade C candidates often require both high-resolution imaging and spectroscopy for confirmation.
Variations of this classification scheme are routinely used in lens finding works \citep[e.g.,][]{2022A&A...668A..73R,2022A&A...666A...1S,2025A&A...697A..14A,Nagam25,Q1-SP048,Q1-SP053}.

\begin{figure*}[t!]
    \centering
    \includegraphics[width=0.381\textwidth]{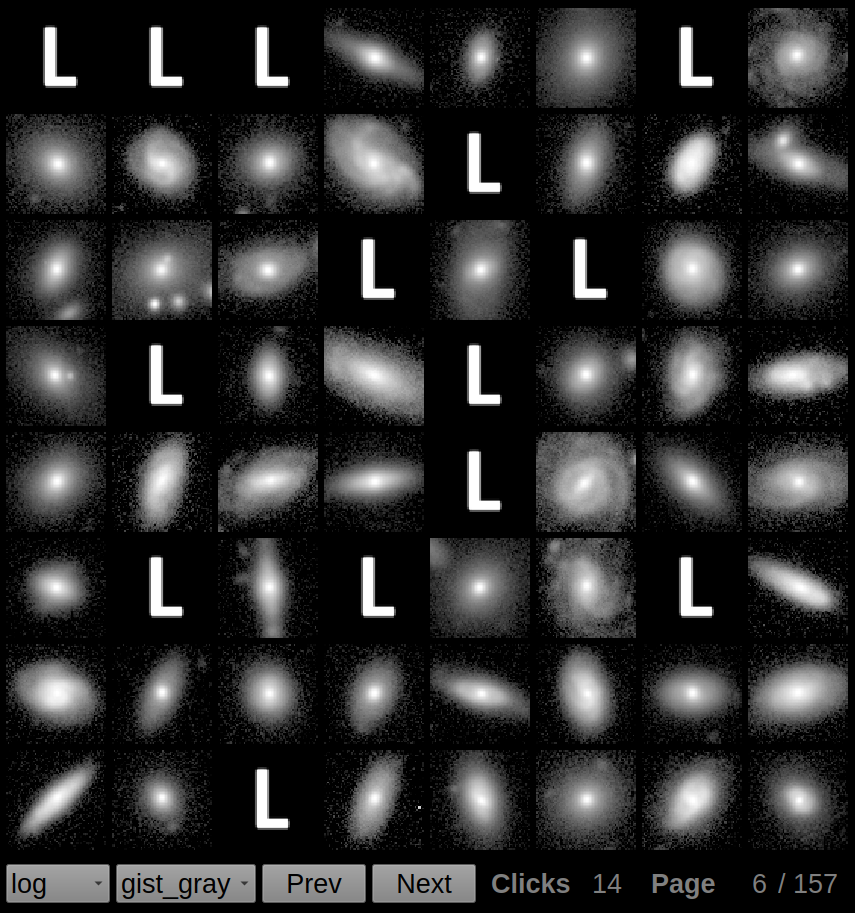}
    \includegraphics[width=0.602\textwidth]{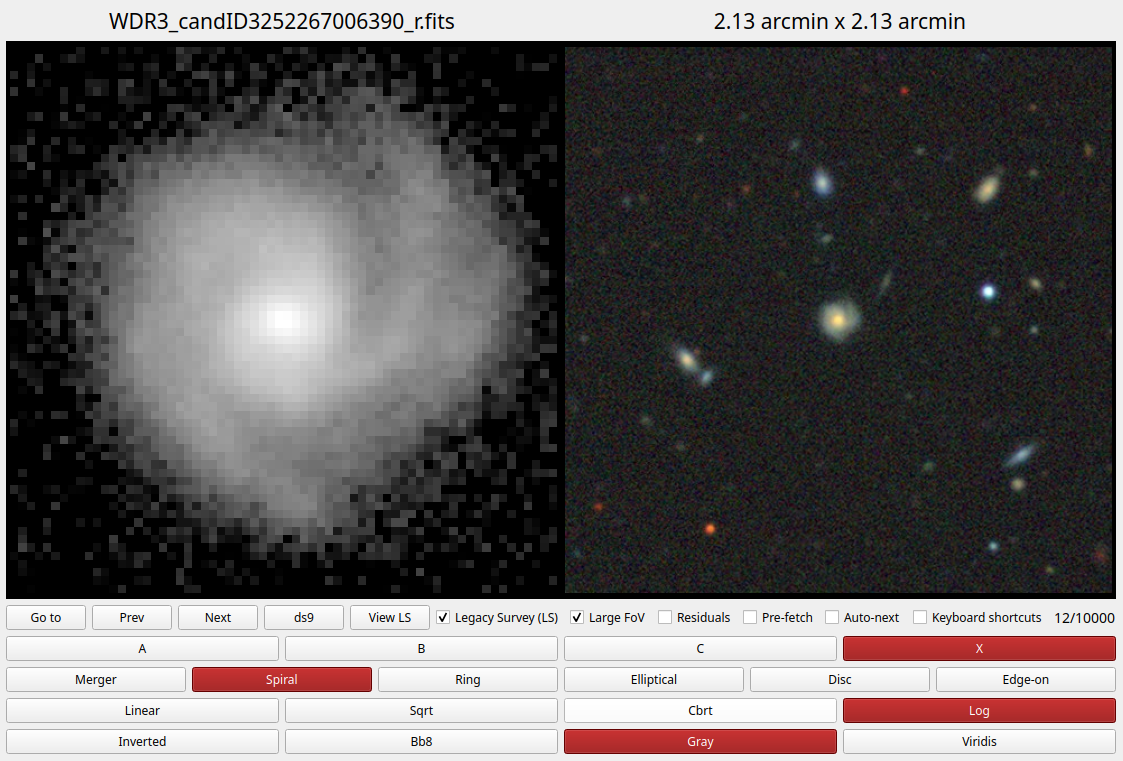}
    \caption{Visualisation tools used for the visual inspections.
    \emph{Left panel}: Mosaic tool showing an $8\times8$ grid whilst highlighting sources classified as `lenses' with a capitalised `L' instead of the \ac{UNIONS} stamp.
    \emph{Right panel}: One-by-one sequential tool showing a spiral galaxy on the left side and the larger field of view from Legacy Survey on the right side.
    The source is classified as a `non-lens' and further sub-classified as a `spiral'.
    Both applications are using logarithmic scale and a grey colour-map for the \ac{UNIONS} stamps.
    }
    \label{fig:visualisation_tool}
\end{figure*}

Additionally, the one-by-one sequential tool offers quick access to three $grz$ colour-composite images from the \ac{DESI} Legacy Imaging Survey \citep[][Legacy Survey henceforth]{2019AJ....157..168D}:
a colour-composite image covering the area of the stamp inspected
$(\ang{;;12.256}\,\times\,\ang{;;12.256})$;
a colour-composite image displaying the environment around the stamp in a larger field of view
$(\ang{;2.13}\,\times\,\ang{;2.13})$;
and a colour-composite covering the same area as the stamp, but displaying the residuals of the inference modelling used to construct the \ac{DESI} source catalogue.
We present an example of the two tools in \cref{fig:visualisation_tool}, for which the one-by-one sequential tool is showing the environment around the inspected stamp.

The first two steps of the visual inspection were performed by J.A.A.B., B.C., and F.C.
Each inspector examined a different set of \num{10000} sources from each magnitude bin during the mosaic step.
Since none of the known edge-on lens candidates have UNIONS/CFIS observations, we used simulated mock lenses as a references (see \cref{sect:simulations,fig:examples_mocks}).
All the candidates selected by any inspector were pooled together for individual one-by-one inspection.
Finally, all candidates that received at least one grade A or grade B vote were reinspected in the final group inspection, which involved six expert classifiers: J.A.A.B., B.C., F.C., C.L., R.G., and K.R.
After inspection and discussion of each candidate, the final grade was decided by a majority vote (i.e. at least four out of six votes).
For this stage, the inspectors had access to $r$- and $u$-band \ac{UNIONS} imaging, as well as the previously described Legacy Survey data.
When available, they also used $grizy$ imaging data from \ac{Pan-STARRS} \citep{2016arXiv161205560C, 2020ApJS..251....7F}, and spectroscopy from \acl{SDSS}\acused{SDSS} \citep[SDSS;][]{2017AJ....154...28B}.

\begin{figure*}[t!]
    \centering
    \includegraphics[width=\textwidth]{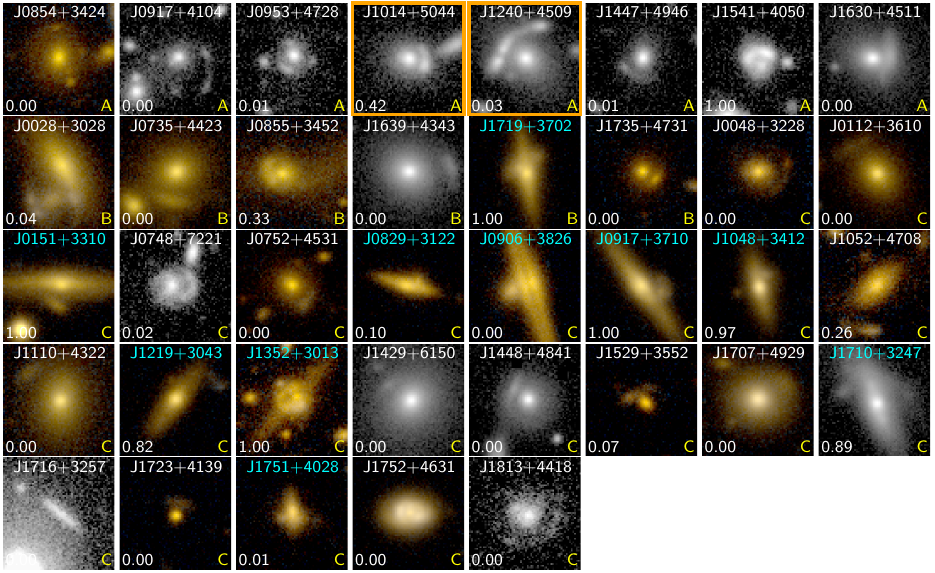}
    \caption{Lens candidates discovered during the prevalence study sorted by visual inspection grade and right ascension.
    We show a colour composite of the $r$- and $u$-bands when the \uband data are available, otherwise we just display a grey scale \rband image.
    The number in the bottom left is the \ac{CNN} score, and the letter in the bottom right corresponds to the final human classification.
    We highlight the edge-on lenses with their name in cyan, and the candidates previously reported in \ac{SLED} with an overlaid orange square.
    }
    \label{fig:candidates_prevalence}
\end{figure*}

The visual inspection rendered eight grade A, six grade B, and 23 grade C candidates.
This includes one grade B edge-on lens candidate and nine grade C.
The candidates are displayed in \cref{fig:candidates_prevalence}.

\subsection{Extrapolating to the whole parent sample}

\begin{table*}[thbp!]
\centering
\caption{Estimate of the number of lens candidates in the parent sample.
}
\smallskip
\label{table:baseline_inspection_results_all_deflectors}
\smallskip
\begin{tabular}{lllllll}
\hline
& & & & & &\\[-9.5pt]
& & & \multicolumn{2}{l}{Edge-on deflector} & \multicolumn{2}{l}{Any deflector} \\
& & & & & &\\[-10pt]
\hline
& & & & & &\\[-9pt]
Grade & Mag. bin & Population size ($N$) & Counts ($k$) & Expected total ($K$) & Counts ($k$) & Expected total ($K$) \\
& & & & & &\\[-10pt]
\hline
& & & & & &\\[-9pt]
A+B & 17 to 18 & \num{328345} & 1 & \phantom{0}11 [8, 35] & 5 & \phantom{0}55 [40, 90] \\
A+B & 18 to 19 & \num{999184} & 0 &  $<\,37$ & 3 & 100 [70, 195] \\
A+B & 19 to 20 & \num{2848367} & 0 &  $<\,108$ & 4 & 380 [271, 677] \\
A+B & 20 to 20.5 & \num{2802381} & 0 &  $<\,106$ & 2 & 187 [129, 431] \\
\hline
& & & & & &\\[-9pt]
C & 17 to 18 & \num{328345} & 5 & \phantom{0}55 [40, 90] & 11 & 120 [95, 166] \\
C & 18 to 19 & \num{999184} & 3 & 100 [70, 195] & 6 & 200 [148, 317] \\
C & 19 to 20 & \num{2848367} & 1 & \phantom{0}95 [68, 311] & 2 & 190 [131, 438] \\
C & 20 to 20.5 & \num{2802381} & 0 &  $<\,106$ & 4 & 374 [267, 666] \\
\hline
& & & & & &\\[-9pt]
\end{tabular}
\tablefoot{
Information derived from the expert visual inspection of $n=\num{30000}$ randomly selected sources per magnitude bin.
Uncertainties are given for a 68\% confidence interval.
Results are shown for both strong lensing in general, and for edge-on lenses only.
If no candidates were found, the estimation corresponds to a prior-dominated upper bound.
The names of the columns in parenthesis match the elements of \cref{eq: prevalence}.
}
\end{table*}
We use Bayes' theorem to estimate the number of lens candidates in the parent sample from the number of candidates found in the visual inspection.
To do so, we made the following assumptions:
(1) We have identified all the lenses in the sample probed (\num{30000} per magnitude bin), and (2)
being randomly selected, the sample probed is representative of the entire parent sample.
With these assumptions, the likelihood $P(k\,|\,K)$ of finding $k$ lens candidates in the probed sample out of a parent sample containing $K$ lens candidates is given by
\begin{equation}
\label{eq: prevalence}
    P(k\,|\,K) = \frac{\binom{K}{k} \binom{N-K}{n-k}}{\binom{N}{n}},
\end{equation}
where $n$ is the number of sources probed and $N$ is the number of sources in the parent sample.
Then, we estimate $P(K\,|\,k)$: the posterior probability that there are $K$ lenses in the parent sample, given that we have found $k$ in the sample probed.
Following Bayes' theorem: $P(K\,|\,k) \propto P(k\,|\,K)P(K)$.
Last, we institute a prior on the number of lenses in the parent sample $P(K)$, such that the probability of a random source showing lensing features ($K/N$) is distributed uniformly logarithmically between $10^{-10}$ and unity.

We made the calculation independently for every magnitude bin, but join the grades A and B to increase the signal.
The results are summarised in \cref{table:baseline_inspection_results_all_deflectors}.
We further classify 10 of the candidates as edge-on based on the morphology of their deflector galaxy.
However, only one of them is deemed grade A, while the rest are deemed grade C.
This good candidate belongs in the brightest magnitude bin, hinting at a core problem of searching for edge-on lenses: the lensed light is very often blended with the deflector, leading to uncertain classifications.
The low signal in our measurement means that the estimates are prior-dominated for all the other magnitude bins and correspond only to upper bounds.
Overall, we estimate at least eight grade A and B edge-on lens candidates, and at least 500 candidates for lensing in general.
This is consistent with a prevalence rate of 1 lens for every \num{10000} sources for lensing in general, and 1 in \num{30000} for edge-on lenses.

\section{Simulating a training dataset}
\label{sect:simulations}
\begin{figure*}[t!]
    \centering
    \includegraphics[width=\textwidth]{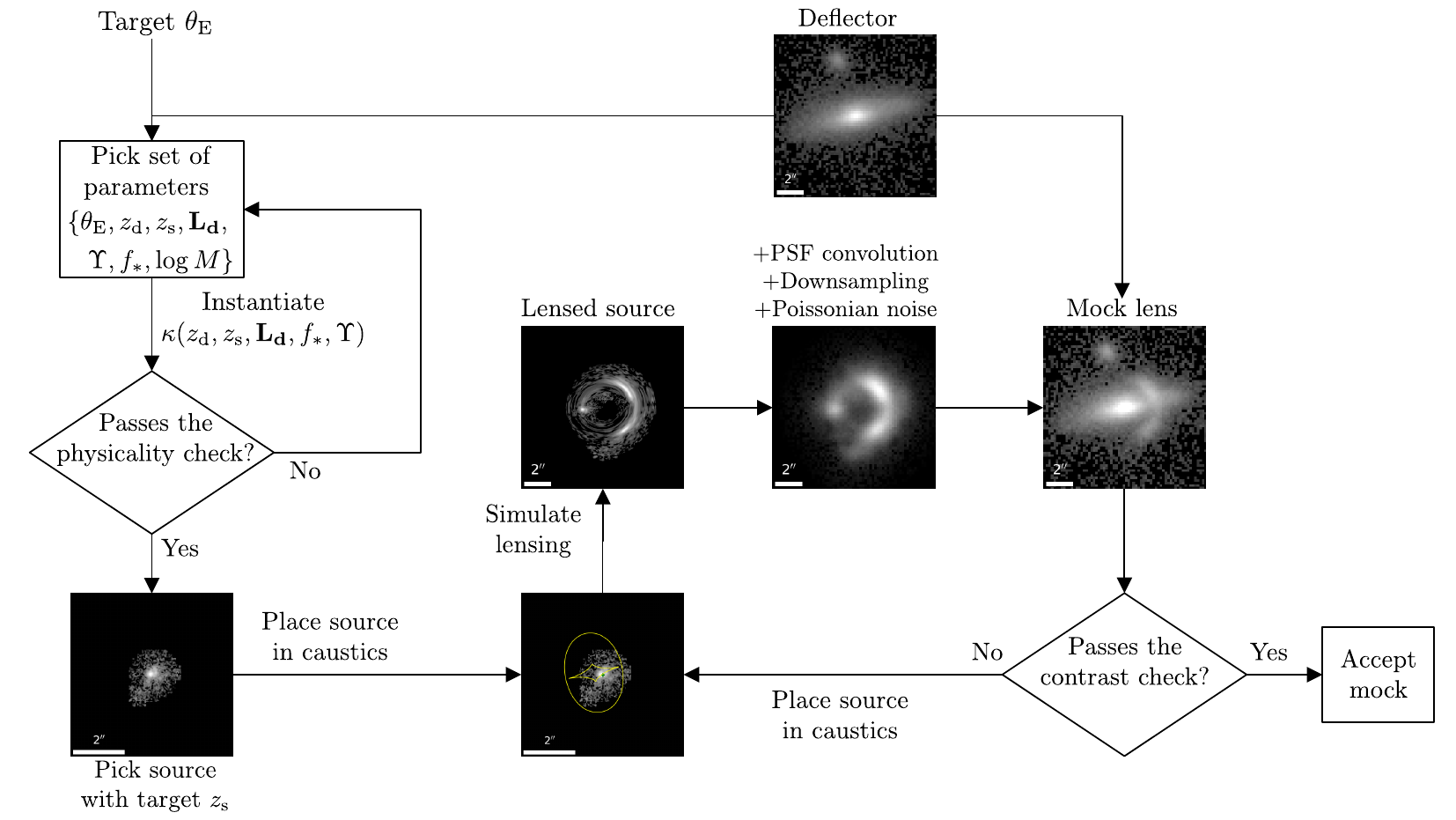}
    \caption{Schematic description of our lens simulation procedure.
    To produce each simulated stamp, we started with a potential deflector and a target Einstein radius, $\theta_{\mathrm{E}}$.
    Then, we picked a set of parameters for a mass model consistent with the target $\theta_{\mathrm{E}}$ and the deflector's light, and we performed a check where we made sure that $\log M$, $f_\ast$, and $\Upsilon$ are within physical ranges.
    After that, we picked a source using the target $z_s$ provided by the mass model parameters, and we placed it randomly around the caustics before simulating its lensing by the mass model.
    Next, we convolved the lensed source image with the deflector's \ac{PSF}, downsampled to match the pixel size of CFIS, and added Poissonian noise.
    At this point, we added the original stamp of the deflector to get the final mock.
    Lastly, we performed a contrast check to make sure that the lensing features are visible.
    If it passed the check, we accepted the mock and proceeded to the next deflector.
    Otherwise, we repositioned the source relative to the caustics and tried a second time.
    }
    \label{fig:simulations}
\end{figure*}

We used the data-driven methodology introduced by \citet{2020A&A...644A.163C} and reproduced by \citet{2022A&A...668A..73R} and \citet{2022A&A...666A...1S} to produce a large dataset of simulated lenses to train our \ac{CNN}.
This required us to select a set of real \ac{UNIONS} sources as potential deflector galaxies.
For each deflector, we assumed a composite mass model comprising both a baryonic component and a dark matter halo.
We then simulated the lensing of real galaxies from deep \textit{HST} imaging data and added the resulting arcs on the \ac{UNIONS} images of our selection of deflectors.
We used the same \textit{HST} sources with \ac{HSC} colours described in \citet{2022A&A...668A..73R}.
For all the lensing calculations and image simulations, we used the Python package \texttt{Lenstronomy} \citep{2018PDU....22..189B,2021JOSS....6.3283B}, and we used \texttt{Astropy} \citep{2022ApJ...935..167A} to handle the imaging data.
\Cref{fig:simulations} gives a visual summary of the complete simulation algorithm.
We describe the different elements of the process in the following sub-sections.

\subsection{Selection of potential CFIS deflectors}
\label{sec:selecting_deflectors}%
We began the process by selecting the edge-on late-type galaxies in our CFIS data.
To do so, we applied an ellipticity cut similar to the one used by \citet{2010A&A...517A..25S}, namely $0.6 < e < 0.92$, with 
\begin{equation}
e \equiv \frac{a^2 - b^2}{a^2 + b^2},
\end{equation}
where $a$ and $b$ are the semi-major and semi-minor axes of the light distribution, respectively.
The lower limit on the ellipticity removes most of the non-edge-on galaxies while the upper limit removes artefacts such as dead columns, stellar spikes, fast asteroids or any other spurious objects with extreme ellipticities.
Due to \ac{PSF} smearing, the percentage of selected galaxies decreases with magnitude, from $18\%$ for the brightest galaxies ($r = 17$) to $12\%$ for the faintest ($r = 20.5$).

Then, we cross-match the selected galaxies with \ac{SDSS} data release 17 \citep{2022ApJS..259...35A} to obtain photometric redshifts (photo-$z$), K-corrections, and distance moduli.
We keep only those galaxies with $\rm{\texttt{photoErrorClass}}\,=\,1$ in the \ac{SDSS} catalogue, ensuring the best possible estimation of the \mbox{photo-$z$} error.
We note that the \ac{SDSS} data do not cover the entire \ac{CFIS} footprint; however, this is not a problem as long as there are enough galaxies with \mbox{photo-$z$} to build a training dataset.
On the other hand, the \mbox{photo-$z$} error increases with magnitude, such that it is about six times larger for the faintest galaxies ($\Delta z \approx 0.13$) than for the brightest ones ($\Delta z \approx 0.02$).
Since the \mbox{photo-$z$} is used to estimate the distance to the deflectors, and from there the luminosity, total mass, and mass-to-light ratio of the mock lenses, a large error in the redshift estimation can introduce significant uncertainties on the physical properties of the mocks.
Fortunately, the relative error in \mbox{photo-$z$} for almost all potential deflectors is between $10\%$ to $30\%$ which is low enough to be acceptable for the generation of mocks.
Thus, only the catastrophic \mbox{photo-$z$} estimations introduce significant errors in the physical parameters of the mock lenses.

Finally, we removed bad deflectors, which we defined as galaxies that, based on their \textit{r}-band image, would require a mass-to-light ratio larger than 12 to generate a lens with Einstein radius of $\theta_{\mathrm{E}} = \ang{;;1.1}$, assuming a source at redshift $z=3.0$ and a \ac{SIE} model for the mass.
We took $\theta_{\mathrm{E}} = \ang{;;1.1}$ because it is almost twice the average seeing, guaranteeing the visibility of the lensing features.
In total, we had \num{649340} galaxies useful as potential deflectors.
For consistency with the K-corrections and \mbox{photo-$z$s}, we used the same cosmology as \ac{SDSS} for all relevant calculations, adopting flat $\Lambda\mathrm{CDM}$ with $\Omega_{\rm m}=0.2739$ and $H_{\rm 0}=\SI{70.5}{\kms\per\mega\parsec}$.

\subsection{Light and mass models}
\label{subsec: light and mass models}
To simulate the mass distribution of deflector galaxies, we used a composite model where baryons and dark matter are treated independently.
This approach is preferable to power-law or isothermal models because, for this type of galaxy, the bulge and disc components {do\-mi\-nate} the lensing mass over the rounder dark halo.
Furthermore, the relative proportion between the bulge, disc, and dark halo has observable effects on the lensing features.

For the baryonic part of the composite model, we used the light profile as seen in the actual data, normalised by a constant stellar mass-to-light ratio, $\Upsilon$.
For the dark halo component, we used a spherical Navarro-Frenk-White profile \citep[NFW, see][]{1996ApJ...462..563N}.
Since the light profiles of the deflectors are highly elliptical, we approximated an elliptical Sérsic profile as the difference between two cored elliptical isothermal profiles.
This is known as a `chameleon' profile, previously used in \citet{2011MNRAS.417.1621D}.
In this work, we followed the parametrisation introduced by \citet{2014ApJ...788L..35S}:
\begin{equation}
    I\left(\Vec{\xi},w_{\rm i},q\right) = \frac{1}{\sqrt{\xi_1^2 + \xi_2^2 / q^2 + 4w_{\rm i}^2/(1 + q)^2}}\,,
\end{equation}
\begin{equation}
    L_{\rm c}\left(\Vec{\xi},w_{\rm c},w_{\rm t},q, L_0\right) = \frac{L_0}{1+q}\left[ I\left(\Vec{\xi},w_{\rm c},q\right)- I\left(\Vec{\xi},w_{\rm t},q\right)\right]\,,
\end{equation}
where $\Vec{\xi} \equiv (\xi_1,\xi_2) = \Vec{\theta} - \Vec{\theta_0}$ is the angular position relative to the centroid of the profile, $q$ is the axis-ratio, and $L_{0}$ is the surface brightness at \ang{;;1} from the centroid along the major axis.
In addition, $w_{\rm c}$ and $w_{\rm t}$ are the sizes of the two cored isothermal profiles that are subtracted from each other, with the additional condition that $w_{\rm t} > w_{\rm c}$.

Because in practice our deflector galaxies can display a prominent bulge component, we also used a `double chameleon' profile to model their light distribution.
The double chameleon is parametrised as the sum of two concentric chameleon profiles, each one with its own ellipticity and core sizes:
\begin{equation}
\resizebox{1\columnwidth}{!}{$
  L_{\rm d}\left(\Vec{\xi},\Vec{X_1},\Vec{X_2},r,L_0\right) =
    L_{\rm c}\left(\Vec{\xi}, \Vec{X_1}, \frac{L_0}{1+r}\right) +
    L_{\rm c}\left(\Vec{\xi}, \Vec{X_2}, \frac{L_0}{1+1/r}\right)
$},
\end{equation}
where the vector $\Vec{X_{\rm i}} \equiv (w_{\rm ic}, w_{\rm it}, q_{\rm i})$ corresponds to the parameters for the single chameleon $i$, and $r$ is the ratio between the profiles along the major axis at \mbox{\ang{;;1}}~away from the common centroid.
We used the chameleon profiles as proxies for the baryonic mass, by replacing the normalisation $L_0$ with $\alpha_1$, where $\alpha_1$ is now the deflection angle along the major axis at $1\arcsec$~away from the centroid.
By contrast, the spherical NFW profile is parametrised in terms of its virial mass $M_{200}$ and concentration $c$.
We used the scaling relations described in \citet{2008MNRAS.390L..64D} to estimate the concentration $c$ for every $M_{200}$.

We fitted the chameleon and double chameleon models to the light of every potential deflector using the \textit{r}-band stamps, their corresponding \ac{PSF} images, and a circular pixel mask with radius \mbox{\ang{;;4.5} around} the centroid of the light distributions.
For this, we used a log-likelihood that combined the standard pixel-wise noise-normalised square error with penalties for ellipticities larger than $e > 0.98$ and for negative fluxes.
We minimised the log-likelihood using first a particle swarm optimiser \citep{488968} with 200 particles and 50 iterations, and then downhill simplex optimisation \citep{10.1093/comjnl/7.4.308} with a maximum of 3000 iterations.
Then, we selected between the chameleon and double chameleon models using the reduced $\chi^2$.
Unsurprisingly, deflectors with prominent bulges were better modelled by the double chameleon, while for bulgeless discs the single chameleon sufficed.
Good models easily reached a reduced $\chi^2$ close 1, since we have reliable estimates of the pixel-wise noise via the \ac{RMS} maps provided by the CFIS reduction pipeline.
Finally, we discarded deflectors with reduced $\chi^2 > 3$.
The whole process left us with a total of \num{558199} potential deflectors with good light models.

\subsection{Reparameterising the mass models}
\label{subsec: reparameterization}
In \texttt{Lenstronomy}, the default parametrisation of mass models is dimensionless.
Instead of modelling lenses in terms of their total mass or mass density, they are parametrised based on the deflection angle caused by their associated convergence map.
This approach allows for the modelling of lens systems while remaining agnostic about the exact cosmology, redshifts, and masses involved.
However, when a specific cosmology and redshifts are considered, the dimensionless convergences can be interpreted as surface mass densities.
This is needed to use scaling relations between the mass and the light distribution of the deflector galaxy.
Since we adopt the same cosmology used in \ac{SDSS}, the total convergence of each potential deflector is completely determined by the redshifts of the deflector and source, $z_{\rm d}$, and $z_{\rm s}$, respectively; as well as by the deflection due to the (Double) Chameleon profile $\alpha_1$, the $M_{200}$ parameter of the NFW, and the light model of the deflector, $\vec{L}$.

However, in order to avoid `unphysical' or unreal-looking lenses, we impose several constraints on the mass models.
More precisely, we force the stellar mass-to-light ratio to be in the range $4.0 < \Upsilon < 9.0$, the fraction of stellar to total mass to be $0.4 < f_\ast < 0.8$ (baryonic fraction), and the total mass to be $10^{10}\,<\,M/M_\sun\,<\,10^{12}$.
All three quantities are calculated in the area enclosed by the tangential critical lines.
In order to do this efficiently we reparameterise the total convergence from $\kappa\left(z_{\rm d},z_{\rm s},\mathbf{L},\alpha_1,M_{200}\right)$ to $\kappa\left(z_{\rm d},z_{\rm s},\mathbf{L}, \Upsilon, f_\ast\right)$.

For every potential deflector, we created a regular grid of parameters: 
\begin{itemize}
    \item $z_{\rm s} \in \{ \mathrm{e}^x \mid x \text{ uniformly sampled between}\ln(z_{\rm d}) \text{ and} \ln(3.0)\}$,
    \item $\Upsilon \in \{4, 5, 6, 7, 8, 9\}$, and
    \item $f_\ast \in \{0.4, 0.48, 0.56, 0.64, 0.72, 0.8\}$.
\end{itemize}
For every combination of parameters in the above grid, $z_{\rm s}, \Upsilon,$ and $f_\ast$; we find the corresponding combination of $z_{\rm s}$, $\alpha_1$, and $M_{200}$; such that $\kappa(z_d,z_s,\vec{L},\Upsilon,f_\ast) = \kappa(z_{\rm d}, z_{\rm s}, \vec{L},\alpha_1,M_{200})$.
Then, we compute $\log(M)$ and $\theta_{\mathrm{E}}$.
For these searches we use the bisection method as is implemented in \texttt{scipy} \citep{2020SciPy-NMeth}.
For each deflector, we save all the compatible values of the seven parameters: $z_{\rm s},\Upsilon,f_\ast,\alpha_1,\log(M_{200}),\log(M)$, and $\theta_{\mathrm{E}}$.
Following that, we estimate the values in-between the grid sampling using the \texttt{scikit-learn} \citep{2011JMLR...12.2825P} implementation of the `Random Sample Consensus' algorithm with a linear regression as base regressor \citep{10.1145/358669.358692}.
We use $\Upsilon$, $f_\ast$, and $\theta_{\mathrm{E}}$ to predict $z_{\rm s}$, $\alpha_1$, $M_{200}$, and $\log(M)$.
For this regression, we normalise the dataset such that each parameter has a mean of 0 and a standard deviation of 1, and generate polynomial features using a polynomial of degree 3.
We take two-thirds of the sampled models for training and leave the remaining one-third for testing.
We keep the interpolated values only if the accuracy on the test set is larger than 98\%.
This has to be done independently for each deflector, since the relationship between the seven parameters depends on the redshift of the deflector and its luminosity.
The output of this procedure is a well populated grid of reasonable mass models, such that for every potential deflector and every combination of physical parameters $\Upsilon$, $f_\ast$, $\log(M)$, and $\theta_{\mathrm{E}}$, we have the corresponding $z_{\rm s}$ and the \texttt{Lenstronomy} parameters $\alpha_1$, and $M_{200}$.

\subsection{Generating a mock lens from a potential deflector}

There is evidence that ML classifiers benefit from a `flat' Einstein radius distribution when training on simulated lenses \citep{2024A&A...692A..72C}.
Consequently, 
we produce the same number of mocks for every target Einstein radius, thus ensuring a flat $\theta_{\mathrm{E}}$ distribution between \mbox{\ang{;;1.2}} and \mbox{\ang{;;2.6}}, with a precision of \mbox{\ang{;;0.1}}.

To produce a single mock lens, we start from a deflector with a good light model, a photometric redshift, and a list of viable mass models.
For simplicity, we ignore the extinction due to the disc, which reduces the apparent luminosity of the deflectors.
We randomly select a mass model that complies with the physical constraints and target $\theta_{\mathrm{E}}$.
This becomes the `target mass model'.
Then, we find a potential source with a redshift equal to the value needed by the target mass model, with a tolerance of $0.02$.
Afterwards, we instantiate the convergence using \texttt{Lentronomy}, and check that the total mass ($M$), the stellar mass-to-light ratio ($\Upsilon$), and the baryonic fraction ($f_\ast$) are consistent with the target mass model.
If any of the parameters is off by more than 25\%, or if the sum of the relative errors is larger than 50\%, we try again with a different target mass model.
If this check fails twice, we skip the deflector and proceed to the next one.
Otherwise, we go on to calculate the caustic lines of the system.

We randomly select between the tangential and radial caustics and place the central pixel of the source inside a polygon with the same shape as the selected caustic, but with a 15\% larger perimeter.
We favour the selection of the tangential over the radial caustic with a probability of 75\% versus 25\%.
Before simulating the lensing of the source by the convergence of the deflector, we interpolate the light of the source from the \textit{HST} image and rotate it by a random angle.
The lensed image is generated with a pixel size exactly 6 times smaller than that of the deflector.
We convolve the high resolution lensed source image with the corresponding \ac{PSF} model of the deflector.
Then, we downsample the resulting image to the CFIS pixel size of \mbox{\ang{;;0.1857}} and add Poissonian noise.
The final mock lens is the sum of the deflector with the simulated image of the lensed source.

At this stage all the simulated lenses are physically motivated images mocking CFIS observations of gravitational lensing.
However, the lensing features can be too faint or too blended with the deflector light, up to the point that it becomes impossible for human experts to discern the lensing features.
This is often the case when simulating small Einstein radius systems and it also affects known lenses when they are seen at worse resolution. 
For example, the SLACS lenses \citep{2006ApJ...638..703B}, which were discovered spectroscopically and thus are rather compact, are not identified as lenses by human experts when seen on \ac{DES} imaging \citep{2023MNRAS.523.4413R}.
Consequently, even if there are arcs or multiple images after simulating the lensing, that by itself is not enough to guarantee that the lensing features are visible and identifiable as such in the final image.
This is especially problematic when simulating lensing by edge-on late-type galaxies because their light profiles tend to be extended and to have substructure, thus increasing the chances of blending between the deflector light and the lensed source.
We address this problem by implementing a `contrast check' to detect if the lensing features are visible in the final image.
If a mock fails the contrast check, we change the position of the source relative to the caustic lines and generate a new mock lens.
If the new mock also fails the contrast check, we reject it and proceed to the next potential deflector.
The details of the contrast check are presented in \cref{appendix:contrast_check}.
Finally, we show a sample of 56 mock lenses with different Einstein radius in \cref{fig:examples_mocks}, and a characterisation of the mock population in \cref{appendix:mock_characterisation}.

\section{Finding lenses with \texorpdfstring{\texttt{CMU\;DeepLens}}{CMU DeepLens}}
\label{sect: lens finding with CMU DeepLens}
\begin{figure*}[htbp!]
    \centering
    \includegraphics[width=0.47\textwidth]{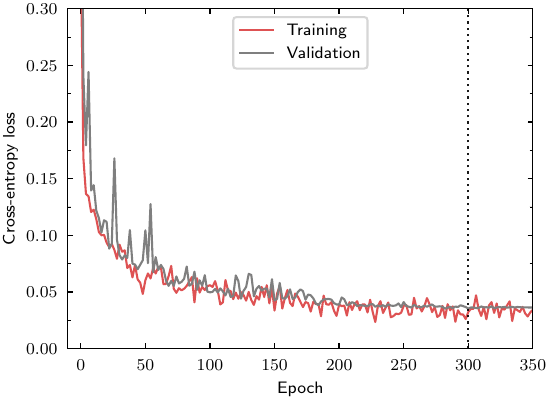}
    \includegraphics[width=0.47\textwidth]{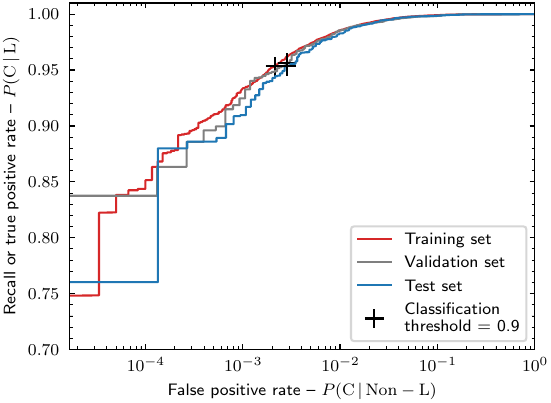}
    
    \caption{Validation metrics for our training of \texttt{CMU\;Deeplens}.
    All the datasets are balanced.
    \emph{Left}: Minimisation of the loss function during training.
    The dotted line marks the epoch with the best validation loss.
    \emph{Right}: \Ac{ROC} curves for the selected model. 
    The area under the curve is 0.9993, 0.9992, and 0.9992 for the training, validation, and test datasets, respectively.
    The `$\times$' marks the values at the classification threshold 0.9.
    }
    \label{fig:loss_and_ROC}
\end{figure*}

\begin{figure*}[htbp!]
    \centering
    \includegraphics[width=0.48\textwidth]{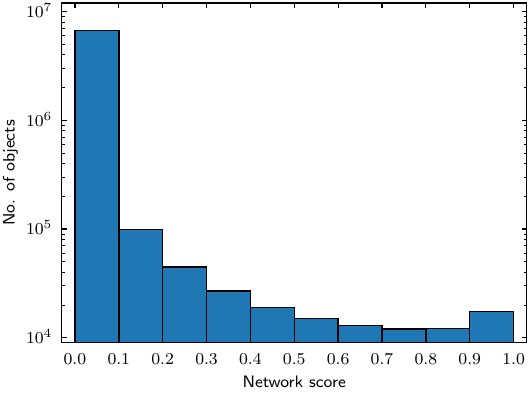}
    \includegraphics[width=0.48\textwidth]{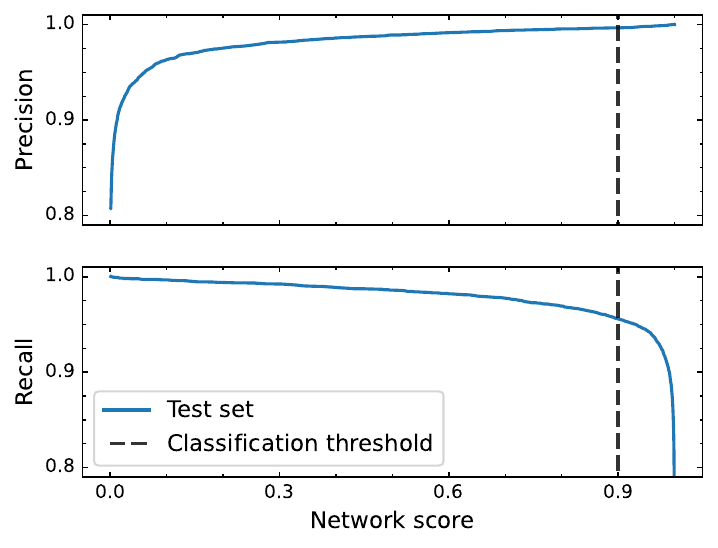}
    \caption{\emph{Left}: Histogram in logarithmic scale of the distribution of network scores for the full parent sample.
    Only \num{17514} of the sources had a score greater than 0.9.
    \emph{Right}: Precision and recall versus network score for the test dataset.
    The dashed line marks the classification threshold applied on the parent sample, 0.9.
    We balanced the precision, recall, and number of sources selected to settle on a classification threshold.
    }
    \label{fig:network_scores}
\end{figure*}

\begin{figure*}[!t]
    \centering
    \includegraphics[width=0.24\textwidth]{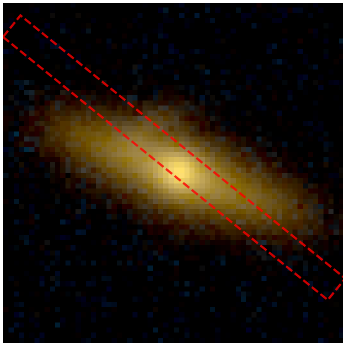}
    \includegraphics[width=0.75\textwidth]{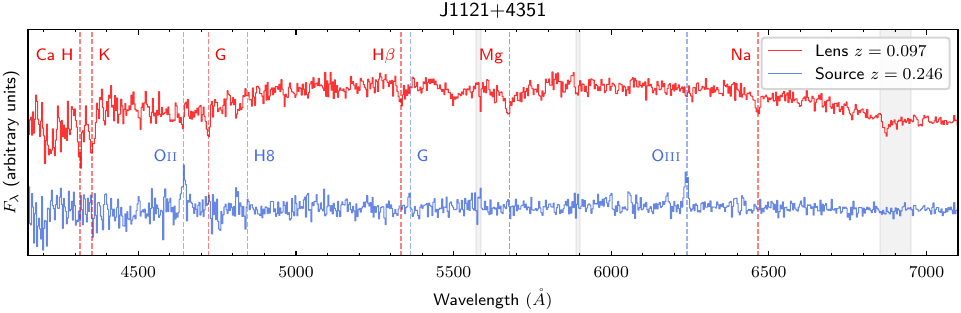}
    \includegraphics[width=0.24\textwidth]{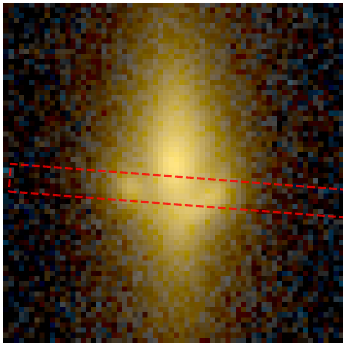}
    \includegraphics[width=0.75\textwidth]{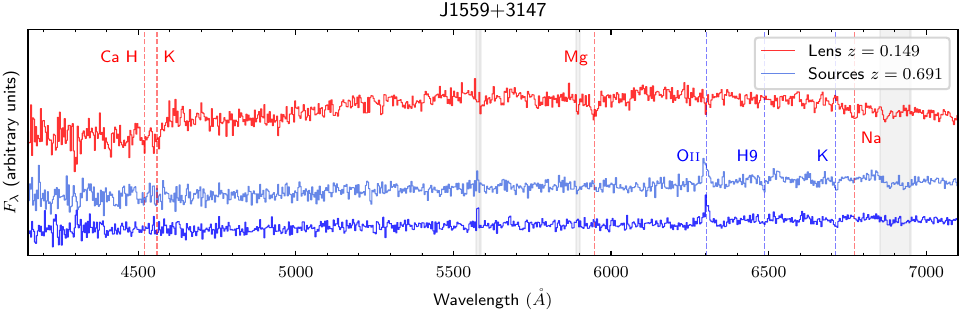}
    \caption{
    Follow-up observations for J1121+4351 and J1559+3147.
    \emph{Left}: UNIONS \rband and \uband imaging of the lens candidates.
    The red dashed line represents the slit position.
    \emph{Right}: Reduced 1D spectra from long-slit spectroscopy taken with the \ac{ALFOSC} instrument in the \ac{NOT}.
    The grey-shaded area marks atmospheric absorption bands.
    Spectral lines are highlighted with vertical dashed lines.
    }
    \label{fig:spectra}
\end{figure*}

We used the \ac{CNN} \texttt{CMU\;DeepLens} \citep{2018MNRAS.473.3895L} for the binary classification task. 
This is a residual neural network \citep[ResNet, see][]{2016cvpr.confE...1H,2016arXiv160305027H} explicitly designed for lens finding.
It is the best performing entry in `The strong gravitational lens finding challenge' \citep{2019A&A...625A.119M} and a modified version has been successfully used in multiple lens finding efforts \citep{2020ApJ...894...78H,2021ApJ...909...27H, 2024ApJS..274...16S, 2022A&A...662A...4S}. 
We reimplemented the network and trained it from scratch using the \texttt{PyTorch} framework \citep{2019arXiv191201703P}.

\subsection{Training}
The dataset consists of \num{150000} cutouts of $66\,\times\,66$ pixels with a pixel size of \mbox{\ang{;;0.1857}}.
The dataset is balanced, with \num{75000} simulated lenses following the prescriptions of \cref{sect:simulations}, and an equal number of non-lenses.
The simulated lenses are distributed uniformly across a range of Einstein radii between \mbox{\ang{;;1.2}} and \mbox{\ang{;;2.6}}, with exactly \num{5000} mocks per \mbox{\ang{;;0.1}} bin.
We present in \cref{fig:examples_mocks} a mosaic with some of the mock lenses in this sample.
For the non-lenses, we reuse the stamps inspected in \cref{sect: estimating the number of lenses (prevalence study)} and keep a flat distribution in magnitude by randomly selecting \num{18750} sources per magnitude bin.
By doing so, we guarantee the absence of potential lens candidates in the negative training set.
We present a characterisation of the image properties of the training data in \cref{appendix:mock_characterisation}.

We split the dataset into three parts: a training set, with 80\% of the sources, and validation and test sets, each with 10\% of the sources.
The subsets are kept balanced, with an equal number of lenses and non-lenses.
During training, each image goes through the following data processing steps:
\begin{enumerate}
    \item Random rotation up to $\ang{180}$  using a bilinear interpolation;
    \item Cropping down to 55 pixels with a new central pixel randomly selected up to 11 pixels away from the previous centre, effectively shifting the image up to \mbox{\ang{;;2}};
    \item Vertical flip with a 50\% chance; and
    \item Normalisation to a mean of zero and a standard deviation of one.
\end{enumerate}
During inference, the images were still cropped to 55 pixels and normalised, but the central pixel remained fixed.

We trained the model by optimising a cross-entropy loss function using a stochastic gradient descent optimiser with an initial learning rate of $10^{-3}$ and a momentum of $0.9$.
The weights of the network are updated using the loss calculated on the training data, while monitoring the loss on the validation set at the end of every training epoch.
If the loss on the validation set does not improve for 20 consecutive epochs, then the learning rate is reduced to half its current value, down to a minimum learning rate of $10^{-8}$, at which the learning rate is kept constant.
The training ends after 50 consecutive epochs without improvement of the validation loss, training for a total for 350 epochs, with the best validation loss being at the 300th epoch.
The batch size during training is 512 and the images are shuffled at every epoch.
Overall, the training takes 17 hours on a NVIDIA GeForce GTX 1080 Ti graphics card.

\subsection{Metrics and inference}
\label{subsect: metrics and inference}
We monitored the evolution of the loss function on both the training and validation datasets.
This is shown in the left panel of \cref{fig:loss_and_ROC}.
The trends of these two sets exhibit no significant divergence, suggesting that overfitting on the training set is not a concern.
In the right panel of \cref{fig:loss_and_ROC}, we present the \ac{ROC} curve for the training, validation, and test sets.
The performance of the algorithm is remarkably consistent across the three datasets, with a marginally better performance on the training set, although, this difference is negligible.
Furthermore, the performance of the trained network on the test and validation datasets is, at first glance, very good, with an area under the \ac{ROC} curve of 0.9992, very close to the optimal 1.0.
However, these metrics underestimate the \ac{FPR} when applied to the parent sample.
Because of the extremely low prevalence of edge-on lenses, estimated to be \mbox{1 in 30\,000} in \cref{sect: estimating the number of lenses (prevalence study)}, the classifier must have a \ac{FPR} of the order of $1/30\,000$ for the sample of CNN-selected lens candidates to have a purity of 50\%.
Failing this, the network will suggest hundreds of false positives per actual edge-on lens candidate.

We applied the trained network to the full parent sample, yielding a network score for each of the \num{6978977} sources.
The distribution of these scores is presented in the left panel of \cref{fig:network_scores}.
We use the distribution of scores, together with the precision and recall curves evaluated on the test set, to choose a reasonable classification threshold for the parent sample.
This is shown in the right panel of \cref{fig:network_scores}.
We select a classification threshold of 0.9, which balances a tolerable number of false positives with the potential number of lenses to be discovered.
In total, the network suggests \num{17514} potential candidates.

Lastly, we used \cref{eq: purity} to estimate the number of good lens candidates selected by the network.
This requires estimates of the model's recall and \ac{FPR}, as well as the general prevalence of edge-on lenses in the parent sample.
The recall of the model is estimated using the mock lenses in the test set, corresponding to $P(\text{C}\, |\, \text{L}) = 0.95$.
Similarly, the \ac{FPR} is estimated from the non-lenses in the test set.
However, this subset contains an equal number of elements in each of the magnitude bins defined in \cref{sect: estimating the number of lenses (prevalence study)}, and thus we calculated the \ac{FPR} per bin.
The weighted average of these values, using the total number of elements per bin in the parent sample as weights, gives the overall \ac{FPR}, corresponding to $P(\text{C}\, |\, \text{Non-L}) = 2.6\times10^{-3}$.
Finally, using the edge-on lens prevalence from \cref{sect: estimating the number of lenses (prevalence study)}, $P(L)\approx1/30\,000$, and \cref{eq: purity}, we predict that $P(\text{L}\, |\, \text{C})\approx1.2\%$ of the candidates selected by the network will be good candidates.
This corresponds to about 208 good edge-on lens candidates in the \num{17514} machine-selected sample.
Since both the \ac{FPR} and the prevalence of edge-on lensing are estimated directly from the data, any deviation from this prediction is most likely due to the estimation of the recall, which is based on mock simulations.

\subsection{Visual inspection of CNN candidates}

Before visually inspecting the sample selected by the CNN, we removed the candidates discovered during the characterisation of the parent sample, that is, the sources presented in \cref{fig:candidates_prevalence}.
We then applied the same three-step visual inspection methodology introduced in \cref{sect: estimating the number of lenses (prevalence study)}.
The first two steps of the visual inspection were performed by J.A.A.B., B.C., and F.C., while the final group inspection was conducted by J.A.A.B., B.C., F.C., C.L., R.G., and
K.R.
After the first two steps, 107 sources were selected for the group inspection.
To that sample, we added lens candidates discovered while prototyping our implementation of \texttt{CMU\;Deeplens}, as well as serendipitous candidates discovered while preparing the simulations used to train the network.
In total, 144 sources were inspected by the six experts in the final one-by-one group inspection.

The final inspection yielded a total of 112 lens candidates, consisting of 27 grade A, 31 grade B, and 54 grade C objects.
Focusing on edge-on lenses, we found 74 candidates: 5 grade A, 19 grade B, and 50 grade C.
Notably, the majority of edge-on lens candidates are grade C, mainly due to the blending of deflector light with lensed images, and the lower lensing cross-section of late-type galaxies compared to \acp{LRG}.
All the edge-on lens candidates are presented in \cref{fig:non_prevalence_mosaic_edge_on}, whilst the non-edge-on candidates are displayed in \cref{fig:non_prevalence_mosaic_not_edge_on}.

\section{NOT spectroscopy}
\label{sect:spectroscopy}

We obtained long-slit spectroscopic follow-up for two lens candidates using the \ac{ALFOSC} instrument at \ac{NOT} over two consecutive nights: 14 and 15 of April, 2023.
For each lens candidate, we acquired two $1200\,\text{s}$ observations using grism 4 with a \mbox{\ang{;;1.0}-wide} slit, providing a spectral resolution of $3.3\,\text{\AA}\,\text{pixel}^{-1}$ and a wavelength coverage spanning $3200\,\text{\AA}$ to $9600\,\text{\AA}$.
The average seeing during the observations was \mbox{\ang{;;0.9}}.

The selected lens candidates are: J1559+3147, a grade A edge-on lens candidate previously reported in \citet{2020A&A...644A.163C} and \citet{2022A&A...666A...1S}; and J1121+4351, a newly discovered grade B edge-on candidate.
Both candidates have existing SDSS spectra, which provide the redshift of the deflectors.
However, although the potential lensing features of J1121+4351 are within the SDSS fibre, the SDSS model residuals lack sufficient \ac{SNR} to constrain the redshift of the source.
By contrast, the brightest lensing features of J1559+3147 are outside the SDSS fibre aperture.
Overall, they are well-suited candidates for confirmation with single-epoch long-slit spectroscopy focused on the lensing features.

We reduced the observations and extracted calibrated 1D spectra following the methodology of \citet{2023MNRAS.520.3305L} and \citet{2024A&A...682A..47D}.
In brief, the 2D spectra were bias-subtracted and cosmic rays masked using a threshold in the absolute Laplacian.
A spatially resolved background was then estimated and removed using the median value around the trace, and an error map was generated by combining background and Poissonian noise.
A wavelength solution was obtained by fitting HeNe arc exposures, and refined by fitting against sky lines. 
The 1D spectra were extracted by dividing the trace into discrete wavelength bins and fitting a Moffat profile to each source in the trace (3 for J1559+3147 and 2 for J1121+4351) at every wavelength bin. 
The parameters of the Moffat profile were then interpolated using a polynomial fit, enabling accurate recovery of the 1D spectra even if the trace shifts spatially.
Finally, the 1D spectra were derived by evaluating the Moffat profiles at every wavelength.
Following this step, we applied flux calibration using the spectrum of the standard star Feige 34.
The resulting 1D spectra are presented in \cref{fig:spectra}.

Regarding the measurement of redshifts, we find for the deflector galaxy in J1121+4351 a series of absorption lines: \mbox{Ca H and K}, Fe G, Mg, and Na, which allows us to confirm the SDSS measurement of $z_{\rm d}=0.097$.
For the potential background source, we detect the emission lines [\ion{O}{II}] and [\ion{O}{III}], indicating star formation and a redshift of $z_{\rm s}=0.246$.
This higher redshift is consistent with the lensing hypothesis, suggesting that the source is a lensed arc.
However, the lack of detection of a counter image, combined with the ground-based resolution of \ac{CFHT}, prevents us from making a definitive confirmation.
Higher resolution imaging is required.
Promisingly, the Euclid Wide Survey will provide high-resolution observations for most of the lens candidates discovered in this work.
This is crucial for edge-on lenses since they have a fair chance of being in a naked-cusp configuration, leading to the absence of a radial counter image and thus requiring high-resolution imaging to better identify tangential lensed features.
In the case of J1559+3147, we observe that the blue spectra share identical features, confirming that they are in fact two images of a single source.
Furthermore, we identify [\ion{O}{ii}] in emission and K in absorption, which we use to obtain a redshift measurement: $z_{\rm s} = 0.691$.
Also, we identify \mbox{Ca H} and K, Mg, and Na in absorption in the spectrum of the deflector, confirming the SDSS redshift: $z_{\rm d}=0.149$.
Remarkably, neither deflector shows emission lines.

\section{Discussion}

\begin{table*}[thbp!]
\centering
\caption{
Summary of the lens search results.
}
\smallskip
\label{table:candidates and sled}
\smallskip
\begin{tabular}{lllll}
\hline
& & & &\\[-9.5pt]
& \multicolumn{2}{l}{Any deflector} & \multicolumn{2}{l}{Edge-on deflector} \\
& & & &\\[-10pt]
\hline
& & & &\\[-9pt]
Grade & Total candidates & New candidates & Total candidates & New candidates \\
& & & &\\[-11pt]
\hline
& & & &\\[-9pt]
A & 35 & 20 & 5 & 4 \\
B & 37 & 36 & 20 & 20 \\
C & 77 & 76 & 59 & 58 \\
All grades & 149 & 132 & 84 & 82 \\
\hline
\end{tabular}
\tablefoot{Includes all candidates regardless of how they were found.
We cross-match our whole sample of lens candidates against the \acf{SLED} to identify the new candidates.
We marginalise by grade, and present results for both strong lensing in general, and edge-on lenses only.}
\end{table*}

\label{sect:discussion}

\begin{table*}[thbp!]
\centering
\caption{
Summary of the performance of the CNN at identifying edge-on lens candidates.
}
\smallskip
\label{table:precision+recal_edge-on}
\smallskip
\begin{tabular}{llllllll}
\hline
& & & & & & &\\[-9.5pt]
Grade & Mag. bin & \specialcell{Candidates\\found} & \specialcell{Candidates\\expected} & Completeness & \specialcell{Purity of the\\ CNN sample} & \specialcell{Prevalence in\\ parent sample} & \specialcell{Ratio of CNN-purity\\ to base prevalence} \\
& & & & & & &\\[-10pt]
\hline
& & & & & & &\\[-9pt]
A+B & 17 to 18 & 4 & 11 & $36\%$ & $7.7\times 10^{-3}$ & $3.3\times 10^{-5}$ & 231.2 \\
A+B & 18 to 19 & 7 & $<37$ & $>19\%$ & $1.9\times 10^{-3}$ & 0 &  \\
A+B & 19 to 20 & 10 & $<108$ & $>9\%$ & $1.1\times 10^{-3}$ & 0 &  \\
A+B & 20 to 20.5 & 0 & $<106$ & 0 & 0 & 0 &  \\
\hline
& & & & &\\[-9pt]
C & 17 to 18 & 9 & 55 & $16\%$ & $1.7\times 10^{-2}$ & $1.7\times 10^{-4}$ & 104.0 \\
C & 18 to 19 & 21 & 100 & $21\%$ & $5.6\times 10^{-3}$ & $1.0\times 10^{-4}$ & 56.2 \\
C & 19 to 20 & 11 & 95 & $12\%$ & $1.2\times 10^{-3}$ & $3.3\times 10^{-5}$ & 35.7 \\
C & 20 to 20.5 & 1 & $<106$ & $>1\%$ & $2.5\times 10^{-4}$ & 0 &  \\
\hline
\end{tabular}
\tablefoot{Metrics are marginalised by grade and magnitude bin to enable direct comparison with \cref{table:baseline_inspection_results_all_deflectors}.
We present the number of candidates found, the expected number of candidates based on \cref{table:baseline_inspection_results_all_deflectors}, and the completeness (number found / expected number).
We also include the fraction of lens candidates among sources inspected for the CNN-selected sample and for the randomly selected sample, respectively; and the ratio between these two, corresponding to the boost in lensing prevalence due to the CNN selection relative to the random selection.}
\end{table*}

\begin{table*}[thbp!]
\centering
\caption{
Summary of the performance of the CNN at identifying lens candidates regardless of deflector type.
}
\smallskip
\label{table:precision+recal_general}
\smallskip
\begin{tabular}{llllllll}
\hline
& & & & & & &\\[-9pt]
Grade & Mag. bin & \specialcell{Candidates\\found} & \specialcell{Candidates\\expected} & Completeness & \specialcell{Purity of the\\ CNN sample} & \specialcell{Prevalence in\\ parent sample} & \specialcell{Ratio of CNN-purity\\ to base prevalence} \\
& & & & & & &\\[-10pt]
\hline
& & & & & & &\\[-9pt]
A+B & 17 to 18 & 4 & 55 & $7\%$ & $7.7\times 10^{-3}$ & $1.7\times 10^{-4}$ & 46.2 \\
A+B & 18 to 19 & 13 & 100 & $13\%$ & $3.5\times 10^{-3}$ & $1.0\times 10^{-4}$ & 34.8 \\
A+B & 19 to 20 & 21 & 380 & $6\%$ & $2.3\times 10^{-3}$ & $1.3\times 10^{-4}$ & 17.0 \\
A+B & 20 to 20.5 & 8 & 187 & $4\%$ & $2.0\times 10^{-3}$ & $6.7\times 10^{-5}$ & 29.9 \\
\hline
& & & & &\\[-9pt]
C & 17 to 18 & 9 & 120 & $8\%$ & $1.7\times 10^{-2}$ & $3.7\times 10^{-4}$ & 47.3 \\
C & 18 to 19 & 22 & 200 & $11\%$ & $5.9\times 10^{-3}$ & $2.0\times 10^{-4}$ & 29.4 \\
C & 19 to 20 & 12 & 190 & $6\%$ & $1.3\times 10^{-3}$ & $6.7\times 10^{-5}$ & 19.5 \\
C & 20 to 20.5 & 1 & 374 & $0\%$ & $2.5\times 10^{-4}$ & $1.3\times 10^{-4}$ & 1.9 \\
\hline
\end{tabular}
\tablefoot{Table follows the layout of \cref{table:precision+recal_edge-on}.}
\end{table*}

\begin{table*}[thbp!]
\centering
\caption{
Overall completeness of the lens search. 
}
\smallskip
\label{table:recall_whole_search}
\smallskip
\begin{tabular}{llllllll}
\hline
& & & & & & &\\[-9pt]
&  & \multicolumn{3}{l}{Edge-on deflector} & \multicolumn{3}{l}{Any deflector} \\
& & & &\\[-10pt]
\hline
& & & &\\[-9pt]
Grade & Mag. bin & \specialcell{Candidates\\found} & \specialcell{Candidates\\expected} & Completeness & \specialcell{Candidates\\found} & \specialcell{Candidates\\expected} & Completeness \\
& & & & & & &\\[-10pt]
\hline
& & & & & & &\\[-9pt]
A+B & 17 to 18 & 7 & 11 & $64\%$ & 12 & 55 & $22\%$ \\
A+B & 18 to 19 & 7 & $<37$ & $>19\%$ & 18 & 100 & $18\%$ \\
A+B & 19 to 20 & 10 & $<108$ & $>9\%$ & 30 & 380 & $8\%$ \\
A+B & 20 to 20.5 & 1 & $<106$ & $>1\%$ & 12 & 187 & $6\%$ \\
\hline
& & & & & & &\\[-9pt]
C & 17 to 18 & 17 & 55 & $31\%$ & 24 & 120 & $20\%$ \\
C & 18 to 19 & 26 & 100 & $26\%$ & 30 & 200 & $15\%$ \\
C & 19 to 20 & 15 & 95 & $16\%$ & 17 & 190 & $9\%$ \\
C & 20 to 20.5 & 1 & $<106$ & $>1\%$ & 6 & 374 & $2\%$ \\
\hline
& & & & & & &\\[-9pt]
\end{tabular}
\tablefoot{
Table includes candidates discovered by the CNN, candidates discovered serendipitously, and candidates discovered during the initial characterisation of the sample.
We use the estimates from \cref{table:baseline_inspection_results_all_deflectors} for the expected total number of candidates.
}
\end{table*}

We cross-matched the coordinates of all the lens candidates against the Strong Lens Database (SLED, Vernardos et al., in prep.) and summarise the results in \cref{table:candidates and sled}.
Overall, out of the 149 candidates identified in this work (grade A, B, and C), only 17 have been reported in \ac{SLED}, including two out of the 82 edge-on lens candidates.
This is consistent with the idea that edge-on lenses are missed in lens searches unless they are explicitly targeted.
On the other hand, given the lack of high-resolution imaging and spectroscopic data, our search is susceptible to contamination by recent mergers and random alignments of galaxies.
But, the \ac{UNIONS} footprint overlaps with that of \Euclid, guaranteeing high-resolution imaging for all of the candidates in the next five years.
This will allow us to confirm or refute many of the candidates, especially in the grade C category.

Regarding our lens finding methodology, specifically the use of a CNN trained and validated on simulations, we note that the validation metrics presented in \cref{subsect: metrics and inference} overestimate the performance of the classifier when applied to the parent sample.
In \cref{subsect: metrics and inference}, we used the prevalence of edge-on lensing as estimated in \cref{sect: estimating the number of lenses (prevalence study)}, along with the recall and \ac{FPR} of the CNN estimated on the test set, to predict that there would be 208 good edge-on lens candidates in the \num{17514} sources selected by the network.
However, only $63$ edge-on candidates ($30\%$ of the predicted number) were found during the visual inspection among the CNN-selected candidates.
This implies that the use of a simulated test set leads to an overestimation of the purity of the network by at least a factor of two.
We believe this is a consequence of imperfect domain adaptation, and that it mainly affects the classifier's recall ($P(\text{C}\, |\, \text{L})$), that is, its ability to identify a real lens as a lens.
Since the classifier is trained on simulations, it is expected that its performance worsens when applied to real lenses.
From comparing the image properties of the mock lenses in the test set against the edge-on candidates found in this work in \cref{appendix:mock_characterisation}, we find that the mock lenses and edge-on candidates both follow similar distributions of background noise, but the edge-on candidates have a considerable higher \ac{SNR}.
Moreover, the simulated mocks followed a flat Einstein radius distribution, which does not reflect the expected distribution for edge-on lenses in CFIS imaging.
Overall, the observed limitations of simulated mocks should be kept in mind when evaluating different networks and classification schemes, which is routinely done using metrics calculated on datasets dominated by mock lenses.

As an alternative to the use of metrics calculated on simulated datasets, we evaluate the effectiveness of our lens search methodology by comparing the CNN-selected sample against the sample that we use in \cref{sect: estimating the number of lenses (prevalence study)} to characterise the parent sample.
In short, we expect the CNN-selected sample to have a higher prevalence of edge-on lenses than the parent sample, while retaining as many edge-on lenses as possible.
We present these metrics in \cref{table:precision+recal_edge-on} for edge-on lenses, and in \cref{table:precision+recal_general} for lensing in general.
We note that using the network increases the prevalence of grade A and B edge-on lensing by up to two orders of magnitude, going from $1/\num{30000}$ to $1/130$ in the brightest magnitude bin, but at the cost of recovering only $36\%$ of the lenses.
However, the situation is less straightforward for the fainter edge-on lenses, since their prevalence is so low that we did not find a single instance of them when characterising the parent sample.
In those cases, our estimates of the completeness are calculated against the upper bounds presented in \cref{table:baseline_inspection_results_all_deflectors}.
Nonetheless, we did find 17 edge-on lens candidates grade A and B with magnitudes between 18 and 20 in the CNN-selected sample, indicating that the CNN is sensitive to them.
We also observe in the last column of \cref{table:precision+recal_edge-on,table:precision+recal_general}, that the improvement in the prevalence decreases with magnitude, which can be explained by the reduced contrast between the lensing features and the deflector's light, leading to noisier classifications by the CNN.
Similarly, the ratio is lower for grade C candidates, being around 100 compared with 230 for the grade A and B.
This suggests that the CNN performs best against clear lens candidates, such as those with higher visual inspection grades and brighter features.

Regarding the objects missed by the CNN, they amount to 4 grade B and 17 grade C edge-on lens candidates.
Notoriously, all of the grade A edge-on lens candidates presented in this work are also part of the CNN-selected sample.
Similarly, there are 12 grade A, 10 grade B, and 16 grade C non-edge-on lens candidates that the CNN missed.
Given a total of 65 non-edge-on lens candidates, the 38 candidates missed by the CNN represent the majority of the sample.
This is expected since the training of the network does not include examples of other types of lenses.

Moreover, we present in \cref{table:recall_whole_search} the completeness of the whole search, including all the candidates regardless of how they were identified.
The search is about $64\%$ complete for the grade A and B edge-on lens candidates in brightest magnitude bin (17 to 18), and at least $19\%$ and $9\%$ complete for those brighter than 19 and 20 magnitudes, respectively.
Regarding the grade C, our search finds $31\%$ of the brightest edge-on lens candidates, and around $20\%$ of those with magnitudes between 18 and 20.
However, even after including the serendipitous discoveries, our search still fails to find a significant number of edge-on lens candidates in the faintest magnitude bin.
Hinting that at magnitudes fainter than 20, the \ac{SNR} falls below the minimum threshold at which we can identify the lensing features.

Finally, having observed the limitations of metrics calculated on simulations, and the fact that the empirical metrics proposed in this work are only available after running the lens search, we propose an iterative approach to improve the purity of lens searches.
In this approach, networks are first trained on simulated mocks plus the few lenses already known in the parent sample.
Then, after visually inspecting the network-selected sample and curating a final list of candidates, the networks should be retrained using the full sample of lens candidates available.
Given that the retraining sample will be on the order of tens of, or at most a few hundred candidates, the retraining must be done carefully, either by freezing most layers and retraining only the final ones or by combining the new lens candidates with the original simulations and gradually removing the simulations from the training data epoch by epoch, leaving only the lens candidates by the end.
Once the model has been retrained, the empirical metrics should be recalculated and used to assess its performance.
Since the prevalence of lensing has already been estimated to evaluate the first model, the calculation of the empirical metrics will be more efficient and less time-consuming for the retrained models.

\section{Conclusions}
\label{sect:conclusions}
We have performed a lens search targeting edge-on late-type galaxies in $\num{3600}\,\rm{deg}^2$ of \ac{UNIONS} \rband data using a \ac{CNN}.
We discovered 4 grade A, 20 grade B, and 58 grade C edge-on lens candidates as well as 16 grade A, 16 grade B, and 18 grade C lens candidates that are not edge-on.
This represents a significant increase in the sample of known edge-on lens candidates, which was previously on the order of one hundred.
Furthermore, we estimate that our new candidates encompass between $9\%$ and $60\%$ of all the discoverable edge-on lens candidates in the parent sample, as determined by the initial inspection of the data in \cref{sect: estimating the number of lenses (prevalence study)}.

We characterised the parent sample by performing a visual inspection of \num{120000} randomly selected sources.
This allowed us to establish a baseline against which we can measure the performance of the CNN in the lens finding task.
Moreover, using Bayes' theorem, we extrapolated the number of candidates found among the randomly selected sources to estimate the total number of lens candidates in the parent sample.
Overall, we estimate that there are at least eight grade A and B edge-on lens candidates and about 250 grade C edge-on lens candidates in the entire parent sample.
This is consistent with a prevalence rate of one edge-on lens candidate for every \num{30000} sources.
Also, for lensing in general, we find the prevalence rate to be consistent with one lens candidate for every \num{10000} sources.

Regarding the performance of the CNN classifier, we show that metrics calculated on simulated datasets significantly overestimate the performance of the network.
Thus, we advise against using metrics calculated on simulations.
As an alternative, we used our initial characterisation of the sample to directly evaluate the performance of our classifier without relying on simulated datasets.
That is, we compared the number of lenses and the purity of the CNN-selected sample to the estimated number of lenses and the prevalence of lensing in the parent sample.
An ideal classifier would select all of the lenses estimated to be in the parent sample and nothing else, resulting in a purity of one and a false-positive rate of zero.
We propose that future lens searches follow a similar methodology when evaluating their classifiers,
reporting both the completeness of their sample and the ratio between the purity of the sample and the estimated prevalence of lensing in the parent sample.
Moreover, we also propose the optimisation of these metrics when retraining models on newly discovered lens candidates.

To create a training set for the CNN, we developed a data-driven methodology that allows us to systematically generate mock observations of edge-on lenses.
Our mock observations use real \rband UNIONS observations for the deflector light and deep \textit{HST} observations for the source light, which improves the realism of the simulated images.
Moreover, the simulated lenses had a realistic mass model, with a stellar-mass-to-light ratio, baryonic fraction, and total mass within the accepted ranges for edge-on lenses, ensuring the simulated lenses are physically plausible.

We spectroscopically followed up on two of our edge-on lens candidates: J1121+4351 and J1559+3147.
For the former, we confirmed the redshift of the deflector to be $z_{\rm d}=0.097$ and measured the redshift of the arc at $z_{\rm s}=0.246$, but given the lack of a clear counter image, high-resolution imaging is still required to confirm its lensing nature.
By contrast, the spectra of the lensed images of J1559+3147 showed identical features, confirming its lensing nature; we measured the redshift of the deflector at $z_{\rm d}=0.149$ and the lensed source at $z_{\rm s}=0.691$.

Finally, we prove the feasibility of finding edge-on lenses in large-area optical surveys using artificial intelligence, as long as the lens search targets edge-on lenses explicitly.
However, even though the \ac{UNIONS} \rband data have an excellent average seeing (\mbox{\ang{;;0.6}}), we hit the limitations of ground-based observations.
The lower lensing cross-section of late-type galaxies, combined with their extended spatial features, results in most lens candidates suffering from blending between the light of the deflector and the lensing features.
This makes it challenging for human experts to identify lens candidates during visual inspection and can also confuse \ac{CNN} classifiers.
This is evident in the decrease in the number of candidates found as the magnitude increases.
Thus, we conclude that to identify the bulk of the late-type galaxy lenses, it is necessary to mine surveys with an even higher resolution.
This is already the case with \ac{HSC}, which was providing most of the new edge-on late-type lens candidates before this work.
However, the most significant step forward will come from \Euclid,
which will observe $\num{14000}\,\deg^2$ with an average resolution of \mbox{\ang{;;0.2}},
combining a large volume probed with superb resolution, thus making it the most promising target for future searches of lensing by edge-on late-type galaxies.

\bibliography{edge_on_late_type_lenses}

\begin{appendix}
\section{Acknowledgements}

\begin{acknowledgements}
We thank Elodie Savary for fruitful discussions in the early stages of this work.
J.~A.~A.~B., B.~C., F.~C., and K.~R. acknowledge support from the Swiss National Science Foundation (SNSF).

Funded by a Royal Society University Research Fellowship. This project has received funding from the European Research Council (ERC) under the European Union’s Horizon 2020 research and innovation programme (LensEra: grant agreement No 945536).

We are honoured and grateful for the opportunity of observing the Universe from Maunakea and Haleakala, which both have cultural, historical and natural significance in Hawaii. This work is based on data obtained as part of the Canada-France Imaging Survey, a CFHT large program of the National Research Council of Canada and the French Centre National de la Recherche Scientifique. Based on observations obtained with MegaPrime/MegaCam, a joint project of CFHT and CEA Saclay, at the Canada-France-Hawaii Telescope (CFHT) which is operated by the National Research Council (NRC) of Canada, the Institut National des Science de l’Univers (INSU) of the Centre National de la Recherche Scientifique (CNRS) of France, and the University of Hawaii. This research used the facilities of the Canadian Astronomy Data Centre operated by the National Research Council of Canada with the support of the Canadian Space Agency. This research is based in part on data collected at Subaru Telescope, which is operated by the National Astronomical Observatory of Japan.
Pan-STARRS is a project of the Institute for Astronomy of the University of Hawaii, and is supported by the NASA SSO Near Earth Observation Program under grants 80NSSC18K0971, NNX14AM74G, NNX12AR65G, NNX13AQ47G, NNX08AR22G, 80NSSC21K1572 and by the State of Hawaii.

Based on observations made with the Nordic Optical Telescope, owned in collaboration by the University of Turku and Aarhus University, and operated jointly by Aarhus University, the University of Turku and the University of Oslo, representing Denmark, Finland and Norway, the University of Iceland and Stockholm University at the Observatorio del Roque de los Muchachos, La Palma, Spain, of the Instituto de Astrofisica de Canarias.

Horizon 2020/2021-2025: This project has received funding from the European Union’s Horizon 2020 research and innovation programme under grant agreement No 101004719 (ORP: OPTICON RadioNet Pilot).

The Legacy Surveys consist of three individual and complementary projects: the Dark Energy Camera Legacy Survey (DECaLS; Proposal ID \#2014B-0404; PIs: David Schlegel and Arjun Dey), the Beijing-Arizona Sky Survey (BASS; NOAO Prop. ID \#2015A-0801; PIs: Zhou Xu and Xiaohui Fan), and the Mayall z-band Legacy Survey (MzLS; Prop. ID \#2016A-0453; PI: Arjun Dey). DECaLS, BASS and MzLS together include data obtained, respectively, at the Blanco telescope, Cerro Tololo Inter-American Observatory, NSF’s NOIRLab; the Bok telescope, Steward Observatory, University of Arizona; and the Mayall telescope, Kitt Peak National Observatory, NOIRLab. Pipeline processing and analyses of the data were supported by NOIRLab and the Lawrence Berkeley National Laboratory (LBNL). The Legacy Surveys project is honoured to be permitted to conduct astronomical research on Iolkam Du’ag (Kitt Peak), a mountain with particular significance to the Tohono O’odham Nation.

NOIRLab is operated by the Association of Universities for Research in Astronomy (AURA) under a cooperative agreement with the National Science Foundation. LBNL is managed by the Regents of the University of California under contract to the U.S. Department of Energy.

This project used data obtained with the Dark Energy Camera (DECam), which was constructed by the Dark Energy Survey (DES) collaboration. Funding for the DES Projects has been provided by the U.S. Department of Energy, the U.S. National Science Foundation, the Ministry of Science and Education of Spain, the Science and Technology Facilities Council of the United Kingdom, the Higher Education Funding Council for England, the National Center for Supercomputing Applications at the University of Illinois at Urbana-Champaign, the Kavli Institute of Cosmological Physics at the University of Chicago, Center for Cosmology and Astro-Particle Physics at the Ohio State University, the Mitchell Institute for Fundamental Physics and Astronomy at Texas A\&M University, Financiadora de Estudos e Projetos, Fundacao Carlos Chagas Filho de Amparo, Financiadora de Estudos e Projetos, Fundacao Carlos Chagas Filho de Amparo a Pesquisa do Estado do Rio de Janeiro, Conselho Nacional de Desenvolvimento Cientifico e Tecnologico and the Ministerio da Ciencia, Tecnologia e Inovacao, the Deutsche Forschungsgemeinschaft and the Collaborating Institutions in the Dark Energy Survey. The Collaborating Institutions are Argonne National Laboratory, the University of California at Santa Cruz, the University of Cambridge, Centro de Investigaciones Energeticas, Medioambientales y Tecnologicas-Madrid, the University of Chicago, University College London, the DES-Brazil Consortium, the University of Edinburgh, the Eidgenossische Technische Hochschule (ETH) Zurich, Fermi National Accelerator Laboratory, the University of Illinois at Urbana-Champaign, the Institut de Ciencies de l’Espai (IEEC/CSIC), the Institut de Fisica d’Altes Energies, Lawrence Berkeley National Laboratory, the Ludwig Maximilians Universitat Munchen and the associated Excellence Cluster Universe, the University of Michigan, NSF’s NOIRLab, the University of Nottingham, the Ohio State University, the University of Pennsylvania, the University of Portsmouth, SLAC National Accelerator Laboratory, Stanford University, the University of Sussex, and Texas A\&M University.

BASS is a key project of the Telescope Access Program (TAP), which has been funded by the National Astronomical Observatories of China, the Chinese Academy of Sciences (the Strategic Priority Research Program “The Emergence of Cosmological Structures” Grant No. XDB09000000), and the Special Fund for Astronomy from the Ministry of Finance. The BASS is also supported by the External Cooperation Program of Chinese Academy of Sciences (Grant No. 114A11KYSB20160057), and Chinese National Natural Science Foundation (Grant No. 12120101003, No. 11433005).

The Legacy Surveys imaging of the DESI footprint is supported by the Director, Office of Science, Office of High Energy Physics of the U.S. Department of Energy under Contract No. DE-AC02-05CH1123, by the National Energy Research Scientific Computing Center, a DOE Office of Science User Facility under the same contract; and by the U.S. National Science Foundation, Division of Astronomical Sciences under Contract No. AST-0950945 to NOAO.

Funding for the Sloan Digital Sky 
Survey IV has been provided by the 
Alfred P. Sloan Foundation, the U.S. 
Department of Energy Office of 
Science, and the Participating 
Institutions. 

SDSS-IV acknowledges support and 
resources from the Center for High 
Performance Computing  at the 
University of Utah. The SDSS 
website is www.sdss4.org.

SDSS-IV is managed by the 
Astrophysical Research Consortium 
for the Participating Institutions 
of the SDSS Collaboration including 
the Brazilian Participation Group, 
the Carnegie Institution for Science, 
Carnegie Mellon University, Center for 
Astrophysics | Harvard \& 
Smithsonian, the Chilean Participation 
Group, the French Participation Group, 
Instituto de Astrof\'isica de 
Canarias, The Johns Hopkins 
University, Kavli Institute for the 
Physics and Mathematics of the 
Universe (IPMU) / University of 
Tokyo, the Korean Participation Group, 
Lawrence Berkeley National Laboratory, 
Leibniz Institut f\"ur Astrophysik 
Potsdam (AIP),  Max-Planck-Institut 
f\"ur Astronomie (MPIA Heidelberg), 
Max-Planck-Institut f\"ur 
Astrophysik (MPA Garching), 
Max-Planck-Institut f\"ur 
Extraterrestrische Physik (MPE), 
National Astronomical Observatories of 
China, New Mexico State University, 
New York University, University of 
Notre Dame, Observat\'ario 
Nacional / MCTI, The Ohio State 
University, Pennsylvania State 
University, Shanghai 
Astronomical Observatory, United 
Kingdom Participation Group, 
Universidad Nacional Aut\'onoma 
de M\'exico, University of Arizona, 
University of Colorado Boulder, 
University of Oxford, University of 
Portsmouth, University of Utah, 
University of Virginia, University 
of Washington, University of 
Wisconsin, Vanderbilt University, 
and Yale University.
\end{acknowledgements}

\onecolumn
\FloatBarrier
\section{Lens candidates as seen by UNIONS}
\begin{figure*}[!htb]
    \centering
    \includegraphics[width=0.965\textwidth]{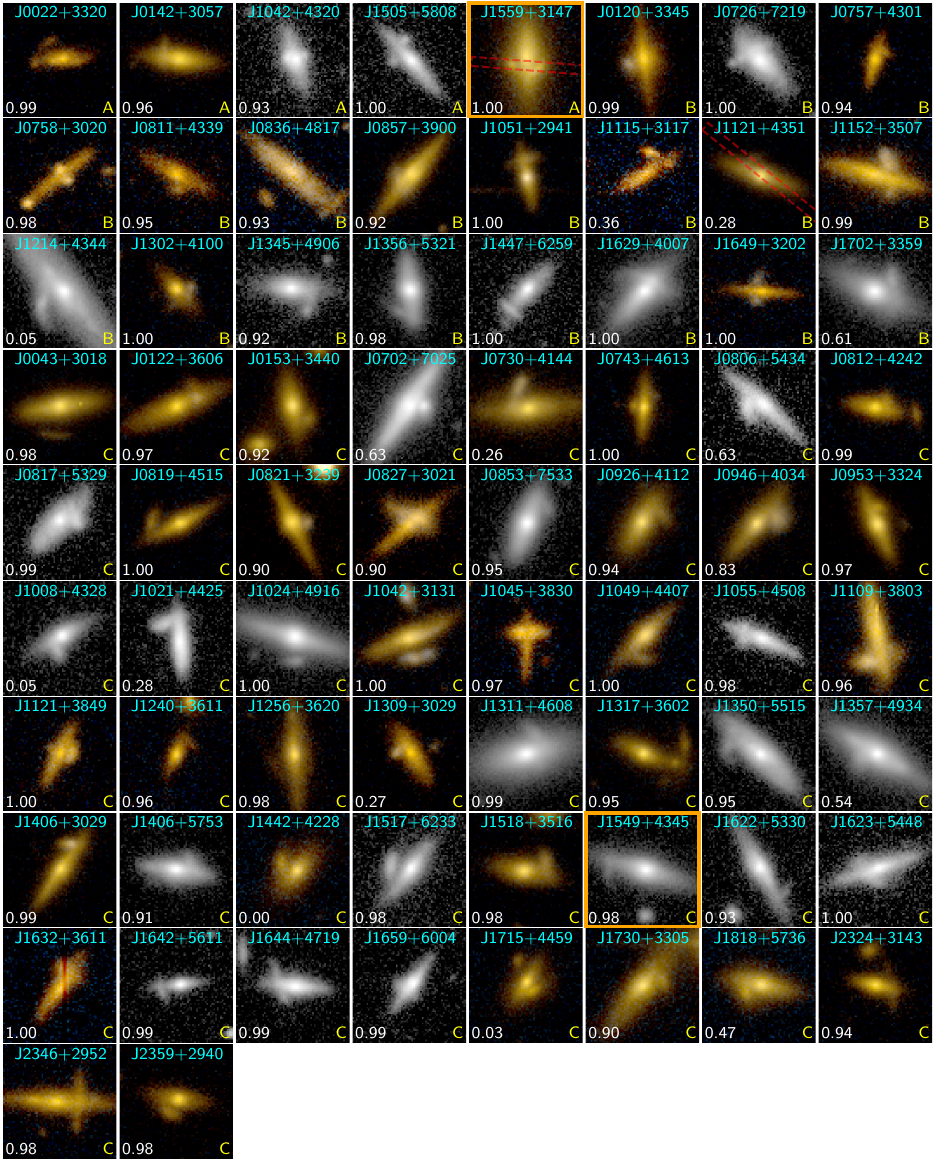}
    \caption{Edge-on lens candidates identified by the trained \ac{CNN} or discovered serendipitously either by prototypes of the classifier or while preparing the mock lenses.
    They correspond to 5 grade A, 19 grade B, and 50 grade C edge-on lens candidates, and of these, only two are already known.
    The layout follows the same scheme from \cref{fig:candidates_prevalence}, with the addition of a rectangle outlined with a red dotted line highlighting the position of the slit for the lens candidates with spectroscopic follow-up.
    }
    \label{fig:non_prevalence_mosaic_edge_on}
\end{figure*}

\begin{figure*}
    \centering
    \includegraphics[width=0.965\textwidth]{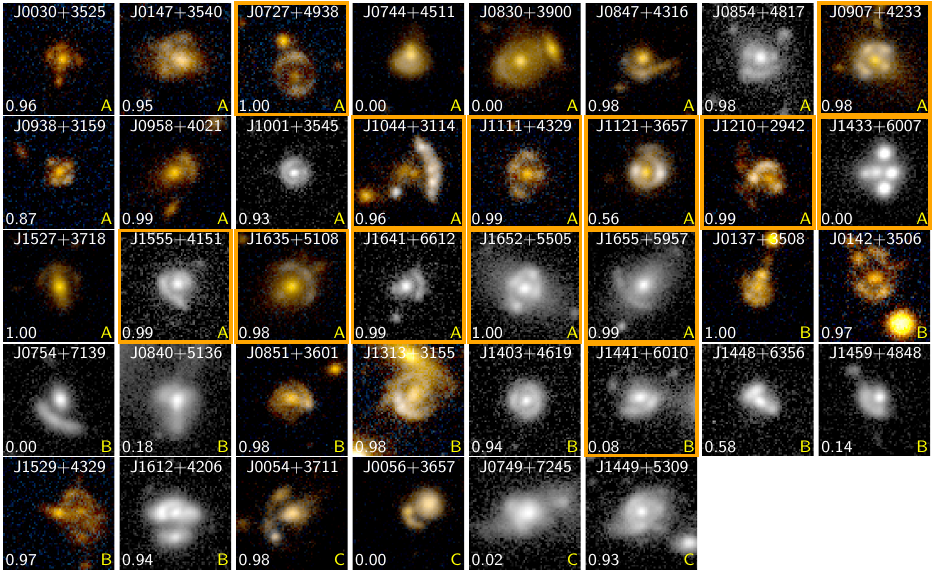}
    \caption{Non-edge-on lens candidates identified by the trained \ac{CNN} or discovered serendipitously either by prototypes of the classifier or while preparing the mock lenses.
    The layout follows the same scheme from \cref{fig:candidates_prevalence}.
    The 38 candidates correspond to 22 grade A, 12 grade B, and 4 grade C.
    Of these, 12 are reported in \ac{SLED}.
    }
    \label{fig:non_prevalence_mosaic_not_edge_on}
\end{figure*}

\FloatBarrier
\twocolumn
\section{The contrast check}
\label{appendix:contrast_check}
\begin{figure}[]
    \centering
    \includegraphics[width=0.43\textwidth]{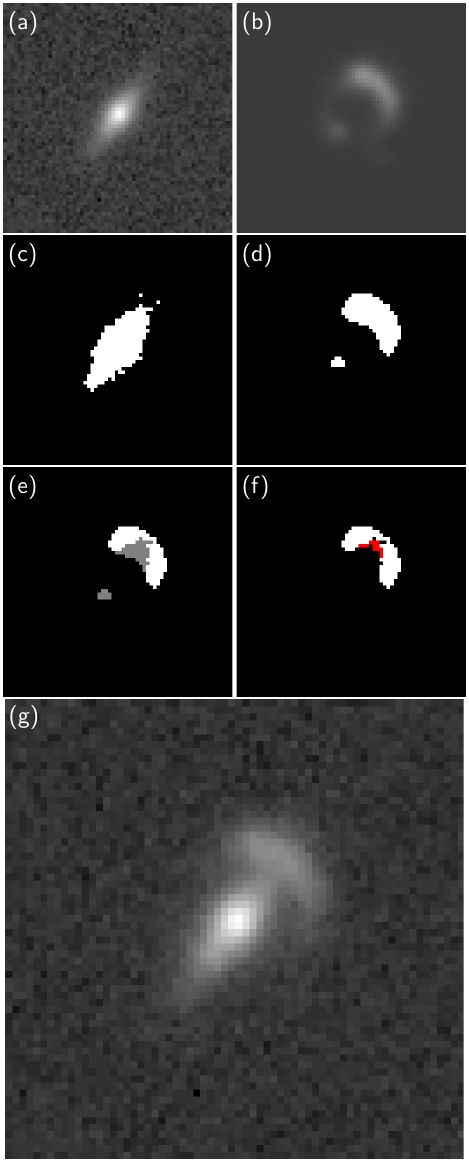}
    \caption{
    Illustration of the contrast check for mock lenses.
    Panels (a) and (b) present the deflector and lensed source light, respectively.
    Panels (c) and (d) show the activated pixels of the deflector and lensed source, respectively.
    Panel (e) presents the activated pixels of the source, but highlighting in grey the pixels that overlap with the deflector mask.
    Panel (f) shows all the pixels dominated by the lensed source light, which are used to decide if the mock passes the contrast check.
    Among them, we highlight in red the pixels that overlap with the deflector, but are still dominated by the lensed source.
    Lastly, panel (g) shows the final mock lens, which passes the contrast check.
    }
    \label{fig:contrast_check}
\end{figure}

In this appendix, we elaborate on the methodology of the contrast check.
We use this check to determine whether the lensing features are visible above the noise background and the deflector galaxy.
Since the images of both the lensed source and the deflector galaxy are sampled on the same pixel grid, we can directly compare corresponding pixels between the two images. 
Thus, the contrast check consists of identifying and counting the pixels that are `dominated by the lensed source light'.
If there are enough such pixels, then the lensing features are deemed as visible and the mock passes the check.
We present the details of the process with the help of \cref{fig:contrast_check}.

\begin{enumerate}[Step 1:]
    \item Estimate the background level $\mu$ and background dispersion $\sigma$ of the deflector image.
    We calculate $\mu$ using the median, and $\sigma$ using the normalised median absolute deviation (MAD).
    \item Create masks for both images by selecting pixels above $\mu + 3\sigma$.
    Panels (a) and (b) show the deflector galaxy and lensed source images, respectively, while panels (c) and (d) display their corresponding masks.
    \item Identify and count the pixels in the lensed source mask that do not overlap with the deflector mask.
    These pixels, shown in white in panel (e), are considered dominated by the lensed source.
    If fewer than five such pixels are found, the contrast check fails.
    \item Identify overlapping pixels between the two masks, which are displayed in grey in panel (e).
    \item Count the overlapping pixels that are dominated by the lensed source light.
    
    \item The total number of pixels dominated by the lensed source light must be larger than 20, and they must amount to at least 30\% of the pixels in the lensed source mask.
    Panel (f) presents all the pixels dominated by the lensed source light.
\end{enumerate}

Finally, we show the final mock lens, which passes the contrast check, in panel (g).

\FloatBarrier
\clearpage

\onecolumn
\section{Examples of mock lenses}

We present in \cref{fig:examples_mocks} examples of the mock lenses that we use to train the \ac{CNN}.
They are grouped by Einstein radius and generated following the methodology described in \cref{sect:simulations}.

\begin{center}
  \begin{minipage}[b]{\textwidth}
    \centering
    \includegraphics[width=\textwidth]{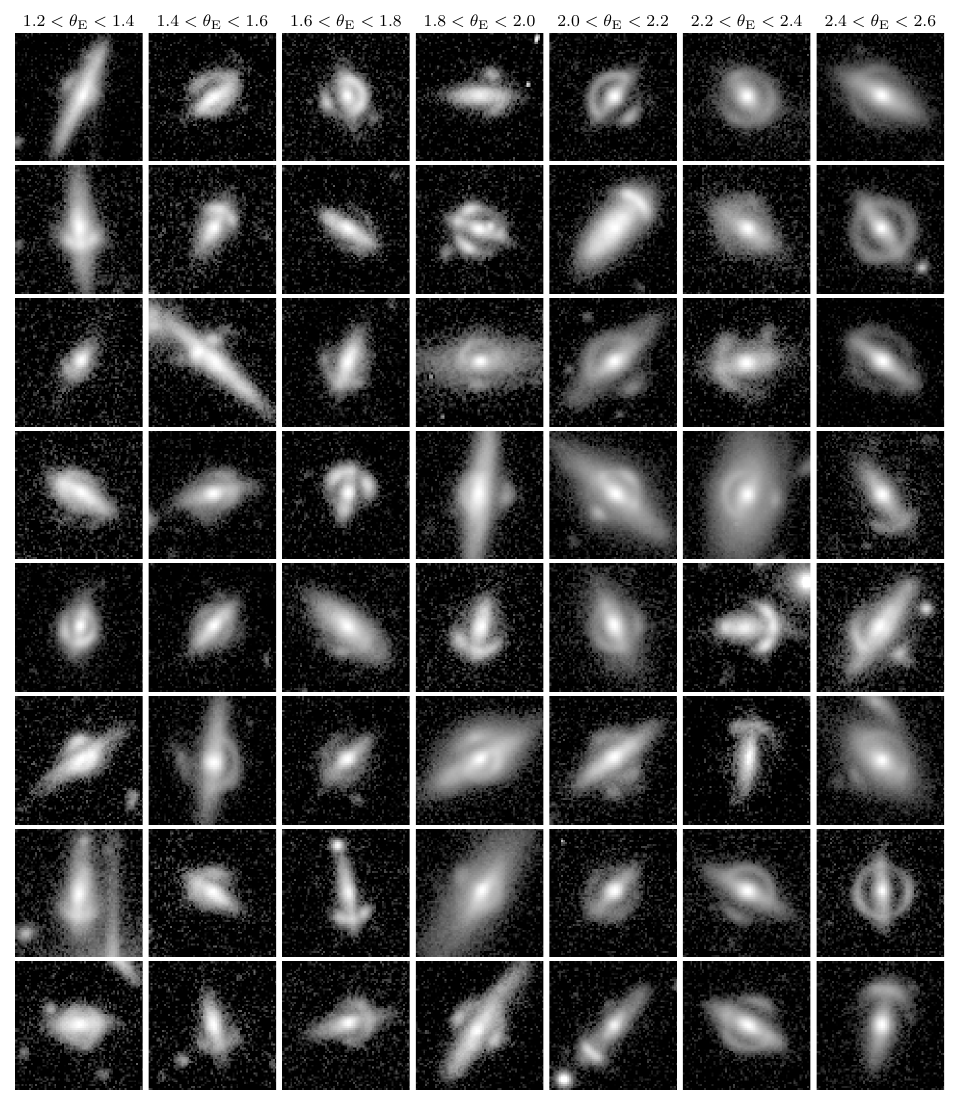}
    \captionof{figure}{
    Mosaic of simulated edge-on lenses mimicking UNIONS \rband observations.
    They follow the prescriptions of \cref{sect:simulations} and are sorted left-to-right by Einstein radius.
    Each image is $\ang{;;12.256}\times\ang{;;12.256}$ in size, corresponding to $66\times66$ pixels.
    }
    \label{fig:examples_mocks}
  \end{minipage}
\end{center}

\FloatBarrier

\newpage
\setlength{\LTcapwidth}{\textwidth}
\twocolumn
\section{Characterisation of the mock population}
\label{appendix:mock_characterisation}

We perform a basic characterisation of the image properties of cutouts of simulated mock lenses compared to those of the newly discovered lens candidates and to the non-lenses used to train the ML model.
In particular, we looked at the background noise of the images and their \ac{SNR}.

To estimate the background noise, we remove the light of the main source using a \ang{;;3} mask centred on the middle of the cutout.
Then, we use 3-sigma clipping to remove any residual light from the main or any other source.
Finally, we estimate the standard deviation of the cutout using the scaled \ac{MAD} of the remaining pixels.
Since the cutouts are background-subtracted, we interpret these standard deviations as the 
average \ac{RMS} of the images.

\begin{figure}[htbp!]
    \centering
    \includegraphics[width=0.5\textwidth]{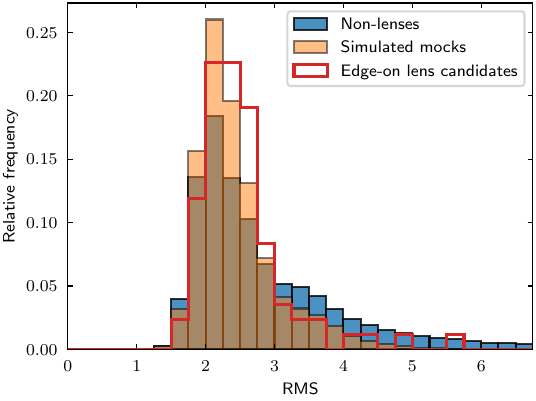}
    \caption{
    \Acf{RMS} distribution of the edge-on lens candidates, and the non-lenses and simulated mocks used to train the \ac{CNN}.
    }
    \label{fig:RMS_hist}
\end{figure}

We show in \cref{fig:RMS_hist} the \ac{RMS} distribution of non-lenses and simulated mocks used to train the network, along with the actual edge-on lens candidates.
We observe that most simulated mocks and lens candidates have \ac{RMS} values between 1.5 and 3, with a tail extending up to 5.
By contrast, non-lenses have a longer and heavier tail extending up to almost 7.
This is consistent with the mock generation methodology described in \cref{sect:simulations}, in which potential deflector galaxies with ``bad'' light models are excluded, thus implicitly removing the noisier data and favouring isolated galaxies.
Moreover, we observe that the edge-on lens candidates follow a similar distribution to that of the simulated mocks, confirming that the mocks are accurate representations in terms of background noise.

\begin{figure}[htbp!]
    \centering
    \includegraphics[width=0.5\textwidth]{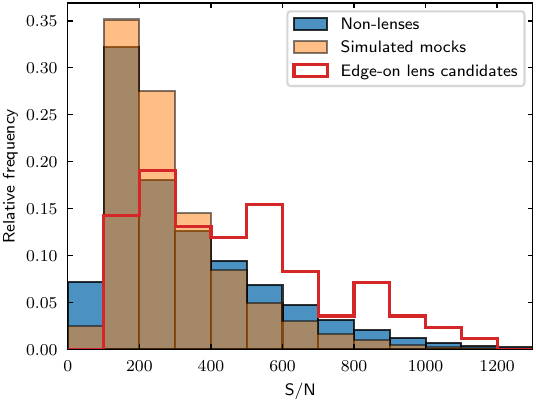}
    \caption{
    \Acf{SNR} distribution of the edge-on lens candidates and the non-lenses and simulated mocks used to train the \ac{CNN}.
    }
    \label{fig:SNR_distribution}
\end{figure}

To estimate the \ac{SNR}, we measure the flux of the deflector using \texttt{SExtractor}.
In order to make the comparison between datasets straightforward, we ignore the previous photometry available in the \ac{CFIS} catalogue, as it used the whole frame instead of small galaxy-centred cutouts and does not include the simulated mocks.
Instead, we run \texttt{SExtractor} on each individual cutout and obtain \texttt{FLUX\_AUTO} and \texttt{FLUXERR\_AUTO} measurements for the deflector galaxy in each.
We present in \cref{fig:SNR_distribution} the distribution of \ac{SNR} for the three datasets.
Since the non-lenses follow a flat magnitude distribution, bright galaxies are over-represented compared to the parent sample.
Thus, they have a heavier tail at higher \ac{SNR}.
On the other hand, the edge-on lens candidates distribution heavily favours higher \ac{SNR} than both the simulated mocks and non-lenses, meaning that the candidates are systematically brighter than the training data.
This difference could help explain the overestimation of the recall when calculated on simulated mocks described in \cref{sect:discussion}, indicating that perhaps we missed dim lens candidates, or that they are simply unnaturally over-represented in the training data.

\newpage
\setlength{\LTcapwidth}{\textwidth}
\onecolumn
\section{Table of edge-on lens candidates}
\begin{longtable}{lllllll} 
\caption{\label{table: edge_on_candidates} List of edge-on lens candidates.
The columns correspond to the \ac{IAU} name of the lens, right ascension, declination, \ac{UNIONS} $u$- and $r$-band photometry using \texttt{SExtractor}'s keyword \texttt{MAG\_AUTO}, the visual inspection grade, and reference to the earliest discovery paper for the previously reported candidates.
} \\
\hline
& & & & & &\\[-9pt]
Name & RA & Dec & \multicolumn{2}{l}{AB magnitude} & Grade & Reference \\ & & & $u$ & $r$ & &\\
& & & & & &\\[-10pt]
\hline
& & & & & &\\[-9pt]
\endfirsthead
\caption{continued.}\\
\hline
& & & & & &\\[-9pt]
Name & RA & Dec & \multicolumn{2}{l}{AB magnitude} & Grade & Reference \\ & & & $u$ & $r$ & &\\
& & & & & &\\[-10pt]
\hline
& & & & & &\\[-9pt]
\endhead
J0022+3320 & \phantom{21}5.7309436 & 33.3474960 &  & $19.9696 \pm 0.0070$ & A &  \\
J0142+3057 & \phantom{1}25.5424303 & 30.9605795 &  & $18.1876 \pm 0.0020$ & A &  \\
J1042+4320 & 160.7169116 & 43.3421165 &  & $19.2626 \pm 0.0053$ & A &  \\
J1505+5808 & 226.3265620 & 58.1382142 &  & $19.7085 \pm 0.0050$ & A &  \\
J1559+3147 & 239.8472065 & 31.7867967 & $19.9230 \pm 0.0249$ & $17.0137 \pm 0.0011$ & A & {\citet{2020A&A...644A.163C}} \\
J0120+3345 & \phantom{1}20.1788702 & 33.7662481 &  & $18.6669 \pm 0.0029$ & B &  \\
J0726+7219 & 111.7042515 & 72.3261701 &  & $18.7084 \pm 0.0029$ & B &  \\
J0757+4301 & 119.4758030 & 43.0255586 &  & $19.8280 \pm 0.0052$ & B &  \\
J0758+3020 & 119.6932428 & 30.3345799 & $22.5128 \pm 0.0925$ & $19.7480 \pm 0.0081$ & B &  \\
J0811+4339 & 122.8543425 & 43.6600055 &  & $19.5493 \pm 0.0079$ & B &  \\
J0836+4817 & 129.1509474 & 48.2978828 &  & $18.8316 \pm 0.0042$ & B &  \\
J0857+3900 & 134.2974125 & 39.0126772 & $20.6838 \pm 0.0317$ & $17.9815 \pm 0.0015$ & B &  \\
J1051+2941 & 162.8862229 & 29.6973580 & $20.6246 \pm 0.0229$ & $18.4459 \pm 0.0021$ & B &  \\
J1115+3117 & 168.7664931 & 31.2907919 &  & $20.2914 \pm 0.0119$ & B &  \\
J1121+4351 & 170.4394276 & 43.8543797 & $20.3017 \pm 0.0210$ & $17.7539 \pm 0.0011$ & B &  \\
J1152+3507 & 178.1576661 & 35.1238394 &  & $18.3464 \pm 0.0026$ & B &  \\
J1214+4344 & 183.5592272 & 43.7499409 &  & $17.3485 \pm 0.0018$ & B &  \\
J1302+4100 & 195.7132585 & 41.0023680 & $22.1846 \pm 0.0832$ & $19.0674 \pm 0.0044$ & B &  \\
J1345+4906 & 206.4890961 & 49.1027004 &  & $19.0217 \pm 0.0043$ & B &  \\
J1356+5321 & 209.1775710 & 53.3648699 &  & $18.1411 \pm 0.0015$ & B &  \\
J1447+6259 & 221.8319148 & 62.9958523 &  & $19.3748 \pm 0.0055$ & B &  \\
J1629+4007 & 247.3489451 & 40.1202681 &  & $17.6530 \pm 0.0015$ & B &  \\
J1649+3202 & 252.3744709 & 32.0443551 &  & $19.6518 \pm 0.0085$ & B &  \\
J1702+3359 & 255.6542281 & 33.9981627 &  & $17.3697 \pm 0.0014$ & B &  \\
J1719+3702 & 259.8905539 & 37.0369870 & $19.9697 \pm 0.0158$ & $17.4750 \pm 0.0012$ & B &  \\
J0043+3018 & \phantom{1}10.9464822 & 30.3134246 & $20.6487 \pm 0.0220$ & $17.4554 \pm 0.0010$ & C &  \\
J0122+3606 & \phantom{1}20.6597358 & 36.1056220 & $21.8596 \pm 0.0576$ & $18.1865 \pm 0.0019$ & C &  \\
J0151+3310 & \phantom{1}27.7945794 & 33.1739183 & $20.0504 \pm 0.0289$ & $17.6152 \pm 0.0017$ & C &  \\
J0153+3440 & \phantom{1}28.2903244 & 34.6712015 & $21.4819 \pm 0.0383$ & $18.2097 \pm 0.0016$ & C &  \\
J0702+7025 & 105.6683506 & 70.4204100 &  & $17.3727 \pm 0.0012$ & C &  \\
J0730+4144 & 112.5626555 & 41.7479087 & $20.4375 \pm 0.0230$ & $17.4948 \pm 0.0018$ & C &  \\
J0743+4613 & 115.8307597 & 46.2167520 & $22.1459 \pm 0.0970$ & $19.0604 \pm 0.0032$ & C &  \\
J0806+5434 & 121.5190539 & 54.5745299 &  & $19.3298 \pm 0.0067$ & C &  \\
J0812+4242 & 123.2002022 & 42.7094191 &  & $18.6894 \pm 0.0024$ & C &  \\
J0817+5329 & 124.4889815 & 53.4859594 &  & $18.8634 \pm 0.0032$ & C &  \\
J0819+4515 & 124.8501970 & 45.2612734 & $21.6504 \pm 0.0742$ & $18.5855 \pm 0.0023$ & C &  \\
J0821+3239 & 125.4266641 & 32.6653315 & $21.7090 \pm 0.0267$ & $18.6794 \pm 0.0029$ & C &  \\
J0827+3021 & 126.7561980 & 30.3640160 & $20.9866 \pm 0.0238$ & $19.0879 \pm 0.0052$ & C &  \\
J0829+3122 & 127.3464117 & 31.3751107 & $21.3560 \pm 0.0391$ & $18.6363 \pm 0.0025$ & C &  \\
J0853+7533 & 133.3933297 & 75.5503074 &  & $17.5901 \pm 0.0012$ & C &  \\
J0906+3826 & 136.5693640 & 38.4449654 & $20.0541 \pm 0.0230$ & $17.1936 \pm 0.0016$ & C &  \\
J0917+3710 & 139.2859126 & 37.1777426 & $19.4074 \pm 0.0142$ & $17.0843 \pm 0.0011$ & C &  \\
J0926+4112 & 141.6601598 & 41.2050782 & $20.4083 \pm 0.0339$ & $17.5684 \pm 0.0012$ & C &  \\
J0946+4034 & 146.7364754 & 40.5765770 & $20.5413 \pm 0.0254$ & $17.9336 \pm 0.0022$ & C &  \\
J0953+3324 & 148.4888058 & 33.4095008 & $21.0903 \pm 0.0204$ & $18.0557 \pm 0.0016$ & C &  \\
J1008+4328 & 152.0535841 & 43.4761871 &  & $18.7321 \pm 0.0028$ & C &  \\
J1021+4425 & 155.3771910 & 44.4282434 &  & $19.0770 \pm 0.0036$ & C &  \\
J1024+4916 & 156.1962473 & 49.2686515 &  & $17.4974 \pm 0.0011$ & C &  \\
J1042+3131 & 160.5660592 & 31.5225667 &  & $18.2438 \pm 0.0021$ & C &  \\
J1045+3830 & 161.4509561 & 38.5006307 &  & $19.4541 \pm 0.0057$ & C &  \\
J1048+3412 & 162.1691522 & 34.2107643 & $19.8473 \pm 0.0116$ & $17.6184 \pm 0.0012$ & C &  \\
J1049+4407 & 162.3353987 & 44.1247252 & $20.3411 \pm 0.0433$ & $18.2829 \pm 0.0019$ & C &  \\
J1055+4508 & 163.7578525 & 45.1374880 &  & $19.7559 \pm 0.0069$ & C &  \\
J1109+3803 & 167.4941749 & 38.0611842 &  & $18.1739 \pm 0.0024$ & C &  \\
J1121+3849 & 170.3750043 & 38.8255148 &  & $19.2811 \pm 0.0057$ & C &  \\
J1219+3043 & 184.8306531 & 30.7172617 & $20.9759 \pm 0.0326$ & $18.1057 \pm 0.0023$ & C &  \\
J1240+3611 & 190.1166129 & 36.1972463 & $23.8831 \pm 0.2015$ & $19.9300 \pm 0.0073$ & C &  \\
J1256+3620 & 194.0349729 & 36.3414725 & $21.2798 \pm 0.0559$ & $18.5258 \pm 0.0035$ & C &  \\
J1309+3029 & 197.2601059 & 30.4984685 &  & $19.3998 \pm 0.0048$ & C &  \\
J1311+4608 & 197.8431059 & 46.1435622 &  & $17.1290 \pm 0.0012$ & C &  \\
J1317+3602 & 199.4890447 & 36.0437890 & $21.6519 \pm 0.0653$ & $18.3090 \pm 0.0021$ & C &  \\
J1350+5515 & 207.5600484 & 55.2558622 &  & $17.9206 \pm 0.0019$ & C &  \\
J1352+3013 & 208.0503035 & 30.2278242 &  & $18.5916 \pm 0.0041$ & C &  \\
J1357+4934 & 209.4420116 & 49.5728381 &  & $17.4802 \pm 0.0010$ & C &  \\
J1406+5753 & 211.5312925 & 57.8922726 &  & $18.5163 \pm 0.0028$ & C &  \\
J1406+3029 & 211.7483549 & 30.4885433 & $21.3309 \pm 0.0480$ & $18.3893 \pm 0.0022$ & C &  \\
J1442+4228 & 220.7377880 & 42.4693250 & $20.9111 \pm 0.0588$ & $18.2496 \pm 0.0024$ & C &  \\
J1517+6233 & 229.4642549 & 62.5608722 &  & $19.1234 \pm 0.0036$ & C &  \\
J1518+3516 & 229.5567630 & 35.2735776 & $21.3932 \pm 0.0417$ & $18.3529 \pm 0.0021$ & C &  \\
J1549+4345 & 237.3001026 & 43.7656831 &  & $18.5109 \pm 0.0022$ & C & {\citet{2022A&A...662A...4S}} \\
J1622+5330 & 245.7261661 & 53.5040794 &  & $18.9184 \pm 0.0037$ & C &  \\
J1623+5448 & 245.9817018 & 54.8155471 &  & $19.1424 \pm 0.0050$ & C &  \\
J1632+3611 & 248.2265743 & 36.1840358 &  & $18.9877 \pm 0.0035$ & C &  \\
J1642+5611 & 250.5932047 & 56.1881179 &  & $20.0957 \pm 0.0076$ & C &  \\
J1644+4719 & 251.1189270 & 47.3251309 &  & $19.4637 \pm 0.0040$ & C &  \\
J1659+6004 & 254.7637000 & 60.0771028 &  & $19.2814 \pm 0.0040$ & C &  \\
J1710+3247 & 257.5724374 & 32.7922491 &  & $17.8119 \pm 0.0018$ & C &  \\
J1715+4459 & 258.8319945 & 44.9888158 & $21.5969 \pm 0.0747$ & $18.1896 \pm 0.0019$ & C &  \\
J1730+3305 & 262.7135754 & 33.0914681 & $20.5849 \pm 0.0467$ & $17.6751 \pm 0.0016$ & C &  \\
J1751+4028 & 267.8865464 & 40.4761924 & $21.2297 \pm 0.0298$ & $19.2625 \pm 0.0043$ & C &  \\
J1818+5736 & 274.7122803 & 57.6039647 & $20.9636 \pm 0.0822$ & $17.8255 \pm 0.0014$ & C &  \\
J2324+3143 & 351.0431226 & 31.7173002 & $22.4031 \pm 0.1010$ & $18.7777 \pm 0.0033$ & C &  \\
J2346+2952 & 356.5326571 & 29.8685802 &  & $18.4314 \pm 0.0032$ & C &  \\
J2359+2940 & 359.8239254 & 29.6824149 & $22.1160 \pm 0.0511$ & $19.0125 \pm 0.0042$ & C &  \\
\hline
& & & & & &\\[-10pt]

\end{longtable}

\newpage
\setlength{\LTcapwidth}{\textwidth}
\onecolumn

\section{Table of other lens candidates}

\begin{longtable}{lllllll} 
\caption{\label{table: non-edge_on_candidates} List of non-edge-on lens candidates.
Same layout as \cref{table: edge_on_candidates}.} \\
\hline
& & & & & &\\[-9pt]
Name & RA & Dec & \multicolumn{2}{l}{AB magnitude} & Grade & Reference \\ & & & $u$ & $r$ & &\\
& & & & & &\\[-10pt]
\hline
& & & & & &\\[-9pt]
\endfirsthead
\caption{continued.}\\
\hline
& & & & & &\\[-9pt]
Name & RA & Dec & \multicolumn{2}{l}{AB magnitude} & Grade & Reference \\ & & & $u$ & $r$ & &\\
& & & & & &\\[-10pt]
\hline
& & & & & &\\[-9pt]
\endhead
J0030+3525 & \phantom{12}7.5491985 & 35.4193203 &  & $20.3806 \pm 0.0145$ & A &  \\
J0147+3540 & \phantom{1}26.7805603 & 35.6831946 & $20.8428 \pm 0.0513$ & $19.4105 \pm 0.0061$ & A &  \\
J0727+4938 & 111.9786573 & 49.6463066 & $21.5481 \pm 0.0983$ & $20.4514 \pm 0.0157$ & A & {\citet{2022ApJ...932..107S}} \\
J0744+4511 & 116.1995056 & 45.1903467 & $20.9530 \pm 0.0288$ & $18.8995 \pm 0.0022$ & A &  \\
J0830+3900 & 127.5749841 & 39.0054697 &  & $18.3616 \pm 0.0028$ & A &  \\
J0847+4316 & 131.9812843 & 43.2683410 & $22.8081 \pm 0.0539$ & $19.9818 \pm 0.0080$ & A &  \\
J0854+3424 & 133.5448406 & 34.4042774 & $23.5610 \pm 0.1858$ & $18.2880 \pm 0.0033$ & A &  \\
J0854+4817 & 133.5597432 & 48.2863870 &  & $19.6335 \pm 0.0083$ & A &  \\
J0907+4233 & 136.8674109 & 42.5504218 &  & $18.5381 \pm 0.0038$ & A & {\citet{2021MNRAS.502.4617T}} \\
J0917+4104 & 139.2739719 & 41.0831402 &  & $20.2640 \pm 0.0132$ & A &  \\
J0938+3159 & 144.5184600 & 31.9889795 &  & $20.4714 \pm 0.0100$ & A &  \\
J0953+4728 & 148.4602610 & 47.4674796 &  & $20.1084 \pm 0.0116$ & A &  \\
J0958+4021 & 149.5197456 & 40.3510648 &  & $19.8450 \pm 0.0090$ & A &  \\
J1001+3545 & 150.3782955 & 35.7634436 &  & $20.2432 \pm 0.0087$ & A &  \\
J1014+5044 & 153.6731083 & 50.7475299 &  & $19.3181 \pm 0.0045$ & A & {\citet{2022A&A...666A...1S}} \\
J1044+3114 & 161.1144384 & 31.2343540 &  & $20.4778 \pm 0.0172$ & A & {\citet{2021ApJ...909...27H}} \\
J1111+4329 & 167.9780720 & 43.4863144 &  & $20.0567 \pm 0.0112$ & A & {\citet{2020A&A...644A.163C}} \\
J1121+3657 & 170.4254175 & 36.9636376 &  & $19.2478 \pm 0.0039$ & A & {\citet{2021ApJ...909...27H}} \\
J1210+2942 & 182.5115475 & 29.7135649 &  & $20.1377 \pm 0.0099$ & A & {\citet{2022ApJ...932..107S}} \\
J1240+4509 & 190.1345492 & 45.1507984 &  & $18.5438 \pm 0.0028$ & A & {\citet{2021ApJ...909...27H}} \\
J1433+6007 & 218.3455098 & 60.1208466 &  & $19.7404 \pm 0.0042$ & A & {\citet{2023A&A...672A.123H}} \\
J1447+4946 & 221.9233871 & 49.7742633 &  & $19.9587 \pm 0.0085$ & A &  \\
J1527+3718 & 231.8380701 & 37.3064436 &  & $19.4510 \pm 0.0041$ & A &  \\
J1541+4050 & 235.4555489 & 40.8415821 &  & $19.3400 \pm 0.0047$ & A &  \\
J1555+4151 & 238.8239469 & 41.8606742 &  & $19.7357 \pm 0.0075$ & A & {\citet{2020A&A...636A..87C}} \\
J1630+4511 & 247.5053897 & 45.1981458 &  & $18.3119 \pm 0.0022$ & A &  \\
J1635+5108 & 248.8455393 & 51.1414327 &  & $18.7642 \pm 0.0039$ & A & {\citet{2022A&A...666A...1S}} \\
J1641+6612 & 250.4827149 & 66.2102977 &  & $19.7497 \pm 0.0114$ & A & {\citet{2021ApJ...909...27H}} \\
J1652+5505 & 253.0912792 & 55.0867535 &  & $18.1866 \pm 0.0033$ & A & {\citet{2021ApJ...909...27H}} \\
J1655+5957 & 253.7973159 & 59.9533299 &  & $18.5849 \pm 0.0049$ & A & {\citet{2024ApJS..274...16S}} \\
J0028+3028 & \phantom{00}7.1125112 & 30.4731739 &  & $17.2964 \pm 0.0012$ & B &  \\
J0137+3508 & \phantom{0}24.3486937 & 35.1433615 &  & $19.8382 \pm 0.0071$ & B &  \\
J0142+3506 & \phantom{0}25.7155714 & 35.1011749 &  & $20.3343 \pm 0.0133$ & B &  \\
J0735+4423 & 113.9150997 & 44.3876939 & $21.0947 \pm 0.0353$ & $17.6047 \pm 0.0019$ & B &  \\
J0754+7139 & 118.6354358 & 71.6619553 &  & $19.5501 \pm 0.0031$ & B &  \\
J0840+5136 & 130.0600175 & 51.6032234 &  & $17.0019 \pm 0.0053$ & B &  \\
J0851+3601 & 132.8226278 & 36.0176713 &  & $19.8241 \pm 0.0064$ & B &  \\
J0855+3452 & 133.9534570 & 34.8735616 &  & $17.6041 \pm 0.0029$ & B &  \\
J1313+3155 & 198.3625467 & 31.9243948 & $19.6699 \pm 0.0266$ & $18.8308 \pm 0.0039$ & B &  \\
J1403+4619 & 210.9584120 & 46.3168808 &  & $19.8105 \pm 0.0058$ & B &  \\
J1441+6010 & 220.3747148 & 60.1690714 &  & $19.3993 \pm 0.0050$ & B & {\citet{2022A&A...666A...1S}} \\
J1448+6356 & 222.1983340 & 63.9471246 &  & $19.0072 \pm 0.0043$ & B &  \\
J1459+4848 & 224.7911651 & 48.8096215 &  & $19.5763 \pm 0.0038$ & B &  \\
J1529+4329 & 232.2841803 & 43.4946833 &  & $20.1853 \pm 0.0089$ & B &  \\
J1612+4206 & 243.0380667 & 42.1025568 &  & $18.9855 \pm 0.0018$ & B &  \\
J1639+4343 & 249.8090159 & 43.7222776 &  & $17.7454 \pm 0.0022$ & B &  \\
J1735+4731 & 263.8469395 & 47.5311135 &  & $19.3923 \pm 0.0052$ & B &  \\
J0048+3228 & \phantom{1}12.0166529 & 32.4692366 & $21.4121 \pm 0.0565$ & $19.2527 \pm 0.0042$ & C &  \\
J0054+3711 & \phantom{1}13.6687485 & 37.1888533 & $20.7709 \pm 0.0189$ & $19.8780 \pm 0.0071$ & C &  \\
J0056+3657 & \phantom{1}14.0748992 & 36.9593161 &  & $20.3934 \pm 0.0063$ & C &  \\
J0112+3610 & \phantom{1}18.0914059 & 36.1756703 & $21.3250 \pm 0.0322$ & $17.7897 \pm 0.0018$ & C &  \\
J0748+7221 & 117.0648816 & 72.3561588 &  & $20.0681 \pm 0.0088$ & C &  \\
J0749+7245 & 117.2753637 & 72.7594174 &  & $17.9379 \pm 0.0027$ & C &  \\
J0752+4531 & 118.1950528 & 45.5174098 & $22.9560 \pm 0.1159$ & $18.9520 \pm 0.0044$ & C &  \\
J1052+4708 & 163.0101111 & 47.1403602 & $21.7768 \pm 0.0623$ & $18.8935 \pm 0.0046$ & C &  \\
J1110+4322 & 167.5611025 & 43.3815384 & $20.2442 \pm 0.0314$ & $17.2265 \pm 0.0016$ & C &  \\
J1429+6150 & 217.4545291 & 61.8346887 &  & $17.4611 \pm 0.0021$ & C &  \\
J1448+4841 & 222.0085402 & 48.6922657 &  & $17.9996 \pm 0.0035$ & C &  \\
J1449+5309 & 222.3743098 & 53.1647651 &  & $18.7549 \pm 0.0035$ & C &  \\
J1529+3552 & 232.2822036 & 35.8812584 &  & $20.1486 \pm 0.0079$ & C &  \\
J1707+4929 & 256.8857557 & 49.4992241 & $19.2383 \pm 0.0079$ & $17.2830 \pm 0.0010$ & C &  \\
J1716+3257 & 259.0916613 & 32.9618078 &  & $18.9359 \pm 0.0096$ & C &  \\
J1723+4139 & 260.8522155 & 41.6502224 & $23.1770 \pm 0.1764$ & $20.2524 \pm 0.0065$ & C &  \\
J1752+4631 & 268.0390198 & 46.5250711 & $18.7596 \pm 0.0033$ & $17.0097 \pm 0.0007$ & C &  \\
J1813+4418 & 273.4938738 & 44.3055507 &  & $20.1083 \pm 0.0142$ & C &  \\
\hline
& & & & & &\\[-10pt]

\end{longtable}

\end{appendix}
\label{LastPage} %
\end{document}